\documentclass[a4paper,fleqn]{cas-sc}

\usepackage{setspace}
\usepackage{xcolor}
\usepackage{lscape} 
\usepackage{natbib}
\usepackage{algorithm2e}
\usepackage{subcaption}
\RestyleAlgo{ruled}
\usepackage{bbm}

\def\tsc#1{\csdef{#1}{\textsc{\lowercase{#1}}\xspace}}
\tsc{WGM}
\tsc{QE}
\tsc{EP}
\tsc{PMS}
\tsc{BEC}
\tsc{DE}

\newcommand{\vect}[1]{\mathbf{#1}}

\newcommand{\addeq}{ \overset{\textrm{\tiny{+}}}{\approx}}

\def\negr#1{\mbox{\boldmath$#1$}}
\def\E{{\mathbb E}} 

\begin{document}
\doublespacing
\let\WriteBookmarks\relax
\def\floatpagepagefraction{1}
\def\textpagefraction{.001}

\shorttitle{Preprint submitted for peer review}

\shortauthors{C. Xian et~al.}

\title [mode = title]{Variational Bayesian analysis of survival data using a log-logistic accelerated failure time model}                      



%
\author[1]{Chengqian Xian}

\cormark[1]


\ead{cxian3@uwo.ca}


\credit{Conceptualization, Methodology, Software, Formal analysis, Writing - Original Draft, Writing - Review \& Editing}

\affiliation[1]{organization={Department of Statistical and Actuarial Sciences, Western University},
    addressline={1151 Richmond Street}, 
    city={London},
    postcode={N6A 5B7}, 
    state={Ontario},
    country={Canada}}

\author[1]{Camila P. E. {de Souza}}
\ead{camila.souza@uwo.ca}
\credit{Conceptualization, Methodology, Writing - Review \& Editing}

\author[1]{Wenqing He}
\ead{whe23@uwo.ca}

\credit{Conceptualization, Methodology, Writing - Review \& Editing}

\author[1, 2]{Felipe F. Rodrigues}
\ead{frodrig7@uwo.ca}
\credit{Conceptualization, Writing - Review \& Editing}

\author[2]{Renfang Tian}
\ead{rtian2@uwo.ca}
\credit{Conceptualization, Writing - Review \& Editing}
\affiliation[2]{organization={School of Management, Economics, and Mathematics, King's University College at Western University},
    addressline={266 Epworth Avenue}, 
    city={London},
    postcode={N6A 2M3}, 
    state={Ontario},
    country={Canada}}

\cortext[cor1]{Corresponding author}



\begin{abstract}
\doublespacing
The log-logistic regression model is one of the most commonly used accelerated failure time (AFT) models in survival analysis, for which statistical inference methods are mainly established under the frequentist framework. Recently, Bayesian inference for log-logistic AFT models using Markov chain Monte Carlo (MCMC) techniques has also been widely developed. In this work, we develop an alternative approach to MCMC methods and infer the parameters of the log-logistic AFT model via a mean-field variational Bayes (VB) algorithm. A piecewise approximation technique is embedded in deriving the VB algorithm to achieve conjugacy. The proposed VB algorithm is evaluated and compared with typical frequentist inferences and MCMC inference using simulated data under various scenarios. A publicly available dataset is employed for illustration. We demonstrate that the proposed VB algorithm can achieve good estimation accuracy and has lower computational cost compared with MCMC methods.
\end{abstract}




\maketitle

\section{Introduction}
\doublespacing
As an alternative to Cox proportional hazards model \citep{Cox_1972}, the accelerated failure time (AFT) model has been widely utilized in survival analysis recently \citep{Webber_2022, Longo_2022, Xu_2022} due to its intuitive interpretation \citep{Wei_1992}. Estimation of parameters and inference under an AFT model are usually likelihood-based  under a frequentist framework \citep{Kalbfleisch_2002, Lawless_2003}. Recent developments have made Bayesian estimation and inference for an AFT model an attractive alternative to likelihood-based methods \citep{Joseph_2001}. Implementations of the AFT model under the framework of Bayesian survival analysis can be found in different scenarios; see, for example, \citet{Lambert_2004, Lesaffre_2008, Zhang_2011} and \citet{Tang_2022}. As for the distributions considered in the parametric AFT model, common choices include log-logistic, Weibull, log-normal, and Gamma distributions. The log-logistic distribution, exhibiting a non-monotonic hazard function, is commonly used in survival analysis when the hazard function presents an inverse U-shape. Empirical analyses in various applications show that the log-logistic distribution is well-suited to model a variety of survival data \citep{Patel_2006, Weng_2014, Thiruvengadam_2021, Rivas_2022}. 

Variational inference (VI), a method developed from machine learning, is used to approximate the posterior distribution of a Bayesian model via optimization \citep{Jordan_1999, Bishop_2006}. \citet{David_2017} presented a comprehensive review of VI from a statistical perspective. As an alternative to Markov Chain Monte Carlo (MCMC) algorithms in Bayesian analysis, the main advantage of VI is its much lower computational cost \citep{David_2017}. In addition, as a Bayesian approach, VI can make use of prior information obtained from similar studies, which are commonly available in survival analysis. Another advantage of VI is that it enables us to conduct inference for small sample sizes since it does not rely on asymptotics \citep{Joseph_2001}, although asymptotic properties for VI methods may still be obtained in some scenarios. For example, \citet{Wang_2019} provided a study on the frequentist consistency of VI when the Kullback–Leibler (KL) minimizer \citep{Kullback_1951} of a normal distribution is considered.

Variational Bayes (VB) is a variational inference method when the KL divergence is used as a criterion to measure the closeness between an approximated posterior density and the exact posterior density in the optimization. VB has been utilized in regression analysis for different statistical problems, such as parametric and nonparametric regression with missing data \citep{Faes_2011}, nonparametric regression with measurement error \citep{Pham_2013}, semiparametric regression for count response \citep{Luts_2015}, high-dimensional linear regression with sparse priors \citep{Ray_2022}, and clustering of functional data via a regression mixture model \citep{xian2022clustering}.

In this paper, we consider the AFT survival model with survival times following a log-logistic distribution and being right censored. We take on a Bayesian approach and develop a VB algorithm to infer the model parameters. To the best of our knowledge, we are the first to build and investigate a VB approach for the AFT survival regression analysis.

The remainder of the paper is organized as follows. Section \ref{sec.background} presents a background of the log-logistic AFT model and the VB inference. We present our methodology including the proposed VB algorithm in Section \ref{sec.method}. In Section \ref{sec.sim}, we conduct simulation studies to evaluate the performance of our method under various scenarios and compare the analysis results with both frequentist analysis and the MCMC analysis. In Section \ref{sec.real}, we apply our proposed method to a real dataset. A discussion on the proposed method is provided in Section \ref{sec.discussion}.

\section{Background}\label{sec.background}

\subsection{Log-logistic accelerated failure time model}

Let $T_i$ be the survival time and $C_i$ be the censoring time of the $i^{th}$ subject in the sample, $i=1, ..., n$. Let $t_i=\min(T_i, C_i)$ and $\delta_i=\mathbbm{1}(T_i \leq C_i)$ be the observed time and the indicator for right censoring of the $i^{th}$ subject, respectively. Then the log-logistic AFT model can be expressed as follows:
\begin{equation}
    \log(T_i)=\vect{X}_i^T\negr{\beta} + bz_i, 
\label{AFT.model}
\end{equation}
where $\vect{X}_i$ is a column vector with length $p, p \geq 2,$ containing $p-1$ fixed effects (covariates) and a constant one to incorporate the intercept (i.e., $\vect{X}_i = (1, x_{i1}, ..., x_{i(p-1)})^T$), $\negr{\beta}$ is the corresponding vector of coefficients for the fixed effects, $z_i$ is a random variable following a standard logistic distribution, and $b$ is a scale parameter. The survival time $T_i$ and censoring time $C_i$ are assumed independent given the covariates $\vect{X}_i$. For the standard logistic distribution, the survival function and density are
\begin{eqnarray}
S_0(z)=\frac{1}{1+e^z},\quad f_0(z)=\frac{e^z}{(1+e^z)^2}, \;-\infty<z<\infty. \nonumber
\end{eqnarray}
Then the log-likelihood for $\negr{\beta}$ and $b$ is
\begin{eqnarray}
l(\negr{\beta}, b)=-r\log b+\sum_{i=1}^{n}\big[\delta_i\log f_0(z_i)+(1-\delta_i)\log S_0(z_i)\big],
\label{eq:loglike}
\end{eqnarray}
where $r=\sum_{i=1}^n\delta_i$ is the number of observed survival times, and $z_i=(y_i-\vect{X}_i^T\negr{\beta})/b, y_i=\log(t_i).$

\subsection{Elements of variational Bayes inference}

In a generic Bayesian model, the posterior density of the parameters is of interest to conduct statistical inference. Consider a Bayesian model with parameter vector $\negr{\theta} \in \Theta$ and observed data $\vect D$. Using the Bayes' theorem, we can obtain the posterior density function by
\begin{equation}
p(\negr{\theta} \vert \vect D)=\frac{p(\negr{\theta}, \vect D)}{p(\vect D)}.
\label{eq:Bayes_theorem}
\end{equation}
However, calculating the posterior density in (\ref{eq:Bayes_theorem}) might not be feasible if there are many parameters and no conjugate prior distributions exist. Therefore, one may alternatively find an approximation to the posterior. While for many years MCMC has stood as the conventional method for attaining this objective, the subsequent paragraphs introduce the elements of variational Bayes inference.

The idea of variational Bayes is to find a variational density $q^*(\negr{\theta})$ from a family of possible densities $Q$ to approximate $p(\negr{\theta} \vert \vect D)$, which can be solved in terms of an optimization problem using the Kullback-Leibler (KL) divergence as a minimization criterion. The KL divergence measures the closeness between the possible densities $q$ in the family $Q$ and the exact posterior density $p$. The KL divergence is defined as 
\begin{eqnarray}
 \text{KL}(q \Vert p)=\E_{q(\negr{\theta})}[\log q(\negr{\theta})] - \E_{q(\negr{\theta})}[\log p(\negr{\theta} \vert \vect D)]. \nonumber
\end{eqnarray}
It can be shown that 
\begin{eqnarray}
\E_{q(\negr{\theta})}[\log q(\negr{\theta})] - \E_{q(\negr{\theta})}[\log p(\negr{\theta} \vert \vect D)]=\int_\Theta q(\negr{\theta})\log\frac{q(\negr{\theta})}{p(\negr{\theta} \vert \vect D)}d\negr{\theta} =\log p(\vect D)-\int_\Theta q(\negr{\theta})\log\frac{p(\negr{\theta}, \vect D)}{q(\negr{\theta})}d\negr{\theta}, \nonumber
\end{eqnarray}
where the last term is the so-called evidence lower bound (ELBO). Since $\log p(\vect D)$ is a constant with respect to $q$, 
\begin{eqnarray}
q^* =  \underset{q \in Q}{\mathrm{argmin}}\,\mbox{KL}(q \Vert p) = \underset{q \in Q}{\mathrm{argmax}}\,\mbox{ELBO}(q).
\label{eq:maxELBO}
\end{eqnarray}
That is, minimizing the KL divergence is equivalent to maximizing the ELBO \citep{Jordan_1999, David_2017}.

The complexity of the variational family, $Q$, determines the complexity of such an optimization problem. It is a great challenge to solve a complex optimization problem corresponding to a complicated variational family. However, when we restrict $Q$ to be the mean-field variational family, $Q_{MF}$, where the parameters and the latent variables are all assumed to be mutually independent and each of them is governed by a distinct factor in the variational density, $q(\negr{\theta})=\prod_{k=1}^K q_k(\theta_k)$ for $q(\negr{\theta})\in Q_{MF}$, the optimization problem in (\ref{eq:maxELBO}) is then changed to 
\begin{eqnarray}
q^*(\negr{\theta}) = \underset{q \in Q_{MF}}{\mathrm{argmax}}\,\mbox{ELBO}(q(\negr{\theta}))=\underset{q \in Q_{MF}}{\mathrm{argmax}}\,\mbox{ELBO}\Big(\prod_{k=1}^K q_k(\theta_k)\Big), 
\label{eq:MFVB}
\end{eqnarray}
where we assume there are $K$ parameters and latent variables, so that $\negr{\theta} =\{\theta_1, ..., \theta_K\}$. 

The coordinate ascent algorithm under the mean-field variational inference \citep{Bishop_2006}, namely coordinate ascent variational inference (CAVI), can be utilized to solve the optimization problem in (\ref{eq:MFVB}). The CAVI algorithm iteratively updates each mean-field variational density factor while keeping the other factors fixed, which makes the variational Bayesian inference a popular alternative to MCMC methods. As shown in \cite{Bishop_2006} and \cite{David_2017}, the update equation for the $k^{th}$ factor ($k=1, ..., K$) in the variational density can be obtained by calculating
\begin{eqnarray}
\log q^*_k(\theta_k) = \E_{-\theta_k} [\log p(\negr{\theta}, \vect D)] + \text{constant},
\end{eqnarray}
where $\log p(\negr{\theta}, \vect D)$ is the log of the joint density of the observed data $\vect D$, the parameters and the latent variables, which is also called the complete-data log-likelihood. The expectation is taken with respect to the variational density of all other parameters and latent variables except the one of interest. The update equation indicates that the expectation on the right-hand side does not involve the $k^{th}$ factor, and therefore can be considered as a coordinate update. With the aid of the CAVI algorithm, the optimization problem (\ref{eq:MFVB}) can be solved by climbing the ELBO to a local optimum \citep{David_2017}.

\section{Methodology}\label{sec.method}
For the log-logistic AFT model specified in (\ref{AFT.model}), we estimate the model parameters, $\negr{\beta}$ and $b$, using a Bayesian framework by further assuming the following prior distributions for $\negr{\beta}$ and $b$: 
\begin{eqnarray}
\negr{\beta} \sim N_p(\negr\mu_0, \sigma_0^2 I_{p\times p}) \; \text{with precision} \; v_0=1/\sigma_0^2, \quad b \sim \text{Inverse-Gamma}\,(\alpha_0, \omega_0), \nonumber
\end{eqnarray}
where $\mu_0, v_0, \alpha_0$ and $\omega_0$ are known hyperparameters \citep{gelman_2004, Faes_2011}.

Our goal is to derive a VB algorithm to approximate $p(\negr{\beta}, b\,\vert\,\vect{D})$, the posterior joint distribution of $\negr{\beta}$ and $b$ given the data $\vect{D}:=\{(t_i, \delta_i, \vect X_i), i=1, ..., n\}$,
with $q^* \in Q_{MF}$ based on the optimization problem specified in (\ref{eq:MFVB}). That is, we assume that $q(\negr{\beta}, b)=q(\negr{\beta})\, q(b)$. The complete-data log-likelihood is then
\begin{eqnarray}
\log p(\vect{D}, \negr{\beta}, b)=\log p(\vect{D}\,\vert\,\negr{\beta},b)+\log p(\negr{\beta})+\log p(b), \nonumber
\end{eqnarray}
where
\begin{eqnarray}
\log p(\vect{D}\,\vert\, \negr{\beta}, b) = -r\log b+\sum_{i=1}^n \Bigg[\delta_i \frac{y_i-\vect{X}_i^T\negr{\beta}}{b}  -(1+\delta_i)\log\Big\{1+\exp\big(\frac{y_i-\vect{X}_i^T \negr{\beta}}{b}\big)\Big\}\Bigg].
\label{eq:loglikelihood}
\end{eqnarray}
By maximizing the ELBO, we have the following solutions \citep{Bishop_2006}:
\begin{eqnarray}
\log q^*(\negr{\beta}) \addeq \E_{q(b)}[\log p(\vect{D}\,\vert\,\negr{\beta},b)+\log p(\negr{\beta})]\quad \text{and} \quad
\log q^*(b) \addeq \E_{q(\negr{\beta})}[\log p(\vect{D}\,\vert\,\negr{\beta},b)+\log p(b)], \nonumber
\end{eqnarray}
where we use $\addeq$ to denote equality up to a constant additive factor for convenience. However, due to the complexity of the logistic distribution and the right censoring scheme, the expectation over the complete-data log-likelihood is challenging to compute. To achieve conjugacy and tractable expectation calculation of $\log p(\vect{D}\,\vert\, \negr{\beta}, b)$ in (\ref{eq:loglikelihood}), we propose piecewise approximations of the function, $f(x) = \log(1+\exp(x)), x \in (-\infty, \infty)$, embedded in deriving the update equations of $q(\negr{\beta})$ and $q(b)$. Illustration of the proposed piecewise approximations is given in Appendix \ref{Appe.B.piece}. 

\subsection{Update equations and the VB algorithm}\label{VBalgo}
The optimal variational densities of $\negr{\beta}$ and $b$, $q^*(\negr{\beta})$ and $q^*(b)$, which are the corresponding approximated posterior distributions, are given as follows:
\begin{eqnarray}
    q^*(\negr{\beta}) \;\text{is a}\; N_p(\negr{\mu}, \Sigma) \; \text{density function, and} \nonumber
\end{eqnarray}
\begin{eqnarray}
    q^*(b) \;\text{is an Inverse-Gamma}(\alpha, \omega) \; \text{density function,} \nonumber
\end{eqnarray}
where the parameters $\negr{\mu}, \Sigma, \alpha$ and $\omega$ are obtained or updated according to Algorithm \ref{VBsurvival} (see derivation details in Appendix \ref{Appe.A.VB}) and $\rho_i, \zeta_i$ and $\varphi_i$ are the piecewise approximation coefficients with formulas provided in Appendix \ref{Appe.B.piece}.

\begin{algorithm}[!ht]
  \KwData{a sample of independent log observed time $y_i$, their corresponding covariate vectors $\vect{X}_i$ and the right censoring indicator $\delta_i, i=1, 2, \cdots, n,$ where $n$ is the sample size; values of hyperparameters: $\negr\mu_0, \sigma_0^2$, $\alpha_0$ and $\omega_0$; convergence threshold $\gamma$ and maximum number of iterations $M$}
  \KwResult{posterior distributions of $\negr\beta$ and $b$ and their parameters: $\Sigma, \negr\mu, \alpha, \omega$}
  \textbf{Initialization}: initialize $\omega=\omega_0$ and $\negr\mu=\negr\mu_0$, set $m=0$ and $ELBO=0$\;
  \textbf{Calculation}: obtain $\alpha$ by $\alpha = \alpha_0 + r$ with $r=\sum_{i=1}^n\delta_i$\;
  \While{iteration $m<M$ and difference of ELBO $>\gamma$}{
    \Repeat{the ELBO converges}{
      $m=m+1$\;
      $\Sigma^{(m)} \leftarrow \bigg[v_0\textbf{I}+2\E_{q(b)}\Big(\frac{1}{b^2}\Big) \sum_{i=1}^n(1+\delta_i)\zeta_i\vect{X}_i\vect{X}_i^T\bigg]^{-1}$ \;
      $\negr\mu^{(m)}\leftarrow \Bigg[\Bigg\{v_0\,\negr{\mu}_0^T + \sum_{i=1}^n \Bigg(\E_{q(b)}\Big(\frac{1}{b}\Big)\Big(-\delta_i+(1+\delta_i)\rho_i\Big)\vect{X}_i^T + 2\E_{q(b)}\Big(\frac{1}{b^2}\Big) (1+\delta_i)y_i \zeta_i \vect{X}_i^T\Bigg)\Bigg\}\,\Sigma^{(m)}\Bigg]^{T}$ \;
      $\omega^{(m)} \leftarrow \omega_0-\sum_{i=1}^n \Big(\delta_i-(1+\delta_i)\varphi_i\Big)\Big(y_i-\vect{X}_i^T\negr\mu^{(m)}\Big)$ \;
      calculate the current ELBO, $\text{ELBO}^{(m)}$ \;
      calculate the difference of ELBO $=\text{ELBO}^{(m)}-\text{ELBO}^{(m-1)}$\;
    }
  }
  \caption{Variational Bayes Inference of Survival Data using a Log-logistic AFT Model}
  \label{VBsurvival}
\end{algorithm}

\subsection{ELBO calculation}

Our goal is to find $q^*(\cdot)$ by maximizing the ELBO. The ELBO is defined as follows:
\begin{eqnarray}
ELBO(q)=\E_{q}[\log p(\vect{D}, \negr{\beta}, b)]-\E_{q}[\log q(\negr{\beta}, b)], \nonumber
\end{eqnarray}
where $\log p(\vect{D}, \negr{\beta}, b)=\log p(\vect{D}\,\vert\, \negr{\beta}, b)+\log p(\negr{\beta})+\log p(b)$ and
$\log q(\negr{\beta}, b)=\log q(\negr{\beta})+\log q(b)$.

Let $\textit{diff}_{\negr{\beta}}=\E_{q}[\log p(\negr{\beta})]-\E_{q}[\log q(\negr{\beta})]$ and 
$\textit{diff}_{b}=\E_{q}[\log p(b)]-\E_{q}[\log q(b)]$, then
\begin{eqnarray}
ELBO(q)=\E_{q}[\log p(\vect{D}\,\vert\, \negr{\beta}, b)]+\textit{diff}_{\negr{\beta}}+\textit{diff}_{b}.
\label{ELBO.cal}
\end{eqnarray}
With some algebraic manipulations (see details in Appendix \ref{Appe.A.VB}), we have
\begin{eqnarray}
\E_{q}[\log p(\vect{D}\,\vert\, \negr{\beta}, b)] \addeq -r\E_{q(b)}\Big(\log b\Big) +\E_{q(b)}\Big(\frac{1}{b}\Big)\sum_{i=1}^n \big(\delta_i -(1+\delta_i)\varphi_i\big)\,\big(y_i-\vect{X}_i^T\negr{\mu}\big),\nonumber
\end{eqnarray}
\begin{eqnarray}
\textit{diff}_{\negr{\beta}} \addeq -\frac{1}{2}v_0[\text{trace}(\Sigma)+(\negr{\mu}-\negr{\mu}_0)^T(\negr{\mu}-\negr{\mu}_0)] + \frac{1}{2}\log (\vert\Sigma\vert), \nonumber
\end{eqnarray}
\begin{eqnarray}
\textit{diff}_{b} \addeq (\alpha-\alpha_0)\E_{q(b)}(\log b) + (\omega-\omega_0)\E_{q(b)}\big(\frac{1}{b}\big) - \alpha\log \omega. \nonumber
\end{eqnarray}

\subsection{Expectations}
In what follows, we calculate the expectations in the update equations in Algorithm \ref{VBalgo} and the ELBO calculations. All the expectations are taken with respect to the approximated variational distributions. Since $q(b)$ is an Inverse-Gamma$(\alpha, \omega)$, we have
\begin{eqnarray}
\E_{q(b)}\Big(\frac{1}{b}\Big)=\frac{\alpha}{\omega},\nonumber
\end{eqnarray}
\begin{eqnarray}
\E_{q(b)}\Big( \frac{1}{b^2} \Big) = \E_{q(b)}\Big[\Big( \frac{1}{b} \Big)^2 \Big] =\text{Var}_{q(b)}\Big[(\frac{1}{b})\Big]+\Big[\E_{q(b)}(\frac{1}{b})\Big]^2 =\frac{\alpha}{\omega^{2}}+\frac{\alpha^{2}}{\omega^{2}}= \frac{\alpha+\alpha^{2}}{\omega^{2}}, \nonumber
\end{eqnarray}
\begin{eqnarray}
\E_{q(b)}(\log b)=\log(\omega)-\Psi(\alpha), \nonumber
\end{eqnarray}
where $\Psi$ is the digamma function defined as $\Psi(x)=\frac{d}{dx}\log \Gamma(x)$.

\section{Simulation studies}\label{sec.sim}
We conduct simulation studies under various scenarios with different sample sizes and censoring percentages to assess the performance of the proposed VB algorithm (i.e., Algorithm \ref{VBsurvival} in Section \ref{VBalgo}).

\subsection{Simulation scenarios and performance metrics}\label{sub.sec.sim.scenarios}
We generate the log of survival time for the $i^{th}$ subject, $\log(T_i), i=1,..., n$, as follows:
\begin{eqnarray}
\log(T_i)=0.5+0.2 x_{i1} + 0.8 x_{i2} + 0.8 z_i, \nonumber
\end{eqnarray}
where $x_{i1}$, $x_{i2}$, and $z_i$ are mutually independently generated with $x_{i1}\sim N(1, 0.2^2)$, $x_{i2}\sim \text{Bernoulli}(0.5)$ and $z_i \sim \text{logistic}(0,1)$. The censoring time for the $i^{th}$ subject, $C_i$, is generated from a uniform distribution, $\text{uniform}(0, u)$, where $u$ is a positive value controlling the percentage of censoring. Then $t_i = \min(T_i, C_i)$ and $\delta_i = \mathbbm{1}(T_i \leq C_i)$. Take $u = 48$ to achieve a 15\% censoring rate and $u = 17$ to achieve a 30\% censoring rate in our simulations.

In the first study, we consider sample sizes of $n=300$ and $n=600$, and varying censoring percentages of 0\%, 15\%, and 30\%. These combinations yield a total of six distinct scenarios. We consider a prior setting with $\negr{\mu}_0 = (0, 0, 0)^T$, $v_0=0.1$, $\alpha_0=11$ and $\omega_0=10$, which indicates no strong prior information on the parameters. The ELBO convergence threshold is set as $0.01$ which is the default recommendation \citep{Yao_2018}, and the maximum number of iterations is $100$. The performance of our VB algorithm are compared against that from the likelihood-based survival regression, \textit{survreg} in the R package \textit{survival} \citep{survival_book, survival_package} and from the MCMC-based algorithm, the Hamiltonian Monte Carlo (HMC) sampling in the R package \textit{rstan} \citep{rstan}.

The second study is designed to assess the performance of the proposed VB algorithms when the sample size is small. When the sample size is small, the likelihood-based estimation methods may fail to achieve satisfactory results. We change the sample size to $n=30$ from $n=300 $ or $600$ in the previous study to evaluate the proposed method for the performance with a small sample size. We also consider a different prior setting with $\negr{\mu}_0 = (0.3, 0.1, 1.0)^T$, $v_0=0.15$, $\alpha_0=11$ and $\omega_0=8$, which indicates partial information about the hyperparameters is known, although they do not precisely match the true parameter values.

We conduct $N=500$ runs (replicates) for each scenario. In each of the 500 replicates, we apply our proposed method to derive an approximate posterior distribution for each parameter. The mean of the posterior distribution serves as our parameter estimate. The empirical bias and sample standard deviation (SD) as well as the empirical mean squared error (MSE) for each estimate are obtained, where
\begin{eqnarray}
\text{MSE}=\frac{\sum_{i=1}^N (\theta_0 -\hat{\theta}_i)^2}{N}, \nonumber
\end{eqnarray}
and $\hat{\theta}_i$ is the estimate of parameter $\theta$ in the $i^{th}$ replicate, and $\theta_0$ is the true value.


In Bayesian statistics, we also assess estimation accuracy by comparing the advertised coverage of approximate credible intervals to their true proposed coverage. We compute 95\% credible intervals for each parameter in 500 replicates. We prefer equal-tailed intervals (ETI) for fixed effects ($\negr{\beta}$) and highest density intervals (HDI) for the scale parameter ($b$) due to the Inverse-Gamma distribution's asymmetry, as suggested by Kruschke (2015). We also calculate the average interval length from these replicates to gauge estimation precision. For comparison, we contrast the empirical credible interval coverage obtained through VB and MCMC with the empirical confidence interval coverage derived from likelihood estimations using the \textit{survreg} method.

High computational cost is a common issue in MCMC-based Bayesian inference algorithms. We compare the performance of our VB algorithm with the MCMC-based HMC algorithm with respect to total run time of 500 replicates. The HMC algorithm in \textit{rstan} \citep{ashraf_2021} is employed to produce four chains with 2000 iterations for each chain. MCMC summaries are based on 4000 MCMC samples after a 1000 sample burn-in for each of the four chains and with the default thinning of 1. Both the VB and HMC algorithms are implemented within R version 4.2.2 on a computer running the Mac OS X operating system with 1.6 GHz CPU and 8 GB RAM.

\begin{landscape}
\begin{table}[h]
\begin{center}
\begin{minipage}{1.35\textwidth}
\caption{\textit{\textbf{Results for the first simulation study:}} A comparison of numerical estimation results including the empirical Bias, sample SD, MSE, coverage rate and average interval length (Avg.L), from our VB method, the \textit{survreg} and MCMC method under different sample sizes ($n$) and censoring percentages ($p$).}
\label{sim.scen.res.num}
\begin{tabular}{cccccccccccccccccc}
\toprule
\multicolumn{3}{c}{} & \multicolumn{5}{c}{VB algorithm} &  \multicolumn{5}{c}{\textit{survreg}} & \multicolumn{5}{c}{MCMC} \\
\cmidrule(rl){4-8} \cmidrule(rl){9-13} \cmidrule(rl){14-18}
$n$  & $p$        &  & Bias       & SD       & MSE& Coverage\footnotemark[1]     & Avg.L      & Bias     & SD      & MSE& Coverage\footnotemark[2]     & Avg.L & Bias     & SD      & MSE& Coverage\footnotemark[1]      & Avg.L    \\ \midrule
\multicolumn{1}{c}{\multirow{12}{*}{$300$}} & \multirow{4}{*}{$0\%$}  & $\beta_0$     &0.017&	0.410&	0.168& 95&	1.59&	0.023&	0.423&	0.179& 94&	1.63 & 0.018&0.426  & 0.182& 95& 1.65\\
\multicolumn{1}{c}{}                          &                       & $\beta_1$     & -0.013&	0.393&	0.154&95&	1.53&	-0.020&	0.405&	0.164 &95&	1.57&-0.017 & 0.412 &0.170 &  95 &1.59 \\
\multicolumn{1}{c}{}                          &                       & $\beta_2$     & -0.002&	0.161&	0.026&94&	0.62&	0.001&	0.161&	0.026&94&	0.63&0.006 &0.161 &0.026& 95 &0.63\\
\multicolumn{1}{c}{}                          &                       & 
$b$ & 0.001&	0.038&	0.001&96&	0.16	&	-0.004&	0.037&	0.001 &95&	0.15 & 0.004&0.037 &0.001 & 96 &0.15\\ \cline{2-18} 
\multicolumn{1}{c}{}                          & \multirow{4}{*}{$15\%$} & $\beta_0$     & 0.011&	0.412&	0.170&95&	1.62&	0.018&	0.426&	0.181&94&	1.65&0.002&0.434&0.188& 96& 1.68 \\
\multicolumn{1}{c}{}                          &                       & $\beta_1$     & -0.008&	0.398&	0.158&95&	1.56&	-0.014&	0.411&	0.169&95&	1.59&0.003&0.419&0.175&95 &1.61\\
\multicolumn{1}{c}{}                          &                       & $\beta_2$     & -0.003&	0.163&	0.027&94&	0.63&	0.001&	0.164&	0.027&94&	0.63&-0.006&0.165&0.027& 95 &0.64 \\
\multicolumn{1}{c}{}                          &                       & 
$b$   & 0.002&	0.041&	0.002&96&	0.18&	-0.004&	0.041&	0.002 &95&	0.16 &0.003&0.040&0.002 &96 &0.16 \\ \cline{2-18} 
\multicolumn{1}{c}{}                          & \multirow{4}{*}{$30\%$} & $\beta_0$     &0.012&	0.421&	0.177&95	&1.65&	0.021&	0.440&	0.194&94&	1.71&0.001&0.448 &0.200&96& 1.74 \\
\multicolumn{1}{c}{}                          &                       & $\beta_1$     &-0.013&	0.404&	0.163&95&	1.60&	-0.017&	0.423&	0.179 &94&	1.65&0.006&0.431&0.186&95 &1.68\\
\multicolumn{1}{c}{}                          &                       & $\beta_2$     &-0.015&	0.165&	0.027&94	&0.65&	-0.003&	0.168&	0.028&93&	0.66&-0.001&0.171&0.029&95& 0.67\\
\multicolumn{1}{c}{}                          &                       & 
$b$   &-0.003&	0.045&	0.002&96&	0.19&	-0.006&	0.045&	0.002&95&	0.18&0.006&0.045& 0.002&95 &0.18\\ \midrule
\multicolumn{1}{c}{\multirow{12}{*}{$600$}} & \multirow{4}{*}{$0\%$}  & $\beta_0$     &-0.015&	0.308&	0.095&93&	1.13&	-0.012&	0.312&	0.097&94&	1.15&-0.010&0.306& 0.094&94 &1.16\\
\multicolumn{1}{c}{}                          &                       & $\beta_1$     & 0.013&	0.299&	0.089&94	&1.09	&	0.010&	0.303&	0.092&94	&1.11&0.009&0.296&0.088&94& 1.11 \\
\multicolumn{1}{c}{}                          &                       & $\beta_2$    &-0.001&	0.113&	0.013&94&	0.44&	-0.001&	0.113&	0.013&95&	0.44&0.002&0.113&0.013 &95&0.44\\
\multicolumn{1}{c}{}                          &                       & 
$b$ &-0.002&	0.028&	0.001&95&	0.12&	-0.003&	0.027&	0.001&94&	0.11&0.002&0.027&0.001&95 &0.11 \\ \cline{2-18} 
\multicolumn{1}{c}{}                          & \multirow{4}{*}{$15\%$} & $\beta_0$     &-0.015&	0.316&	0.100&94	&1.15	&	-0.011&	0.321&	0.103&93&	1.17&0.017&0.307&0.095&96& 1.18\\
\multicolumn{1}{c}{}                          &                       & $\beta_1$     & 0.011&	0.306&	0.094&94	&1.11&	0.008&	0.311&	0.097&93&	1.13&-0.009&0.301&0.092&96&1.14 \\
\multicolumn{1}{c}{}                          &                       & $\beta_2$     & -0.001&	0.114&	0.013&95&	0.45&	0.001&	0.114&	0.013&95	&0.45&-0.004&0.113&0.013&95&0.45\\
\multicolumn{1}{c}{}                          &                       & 
$b$ &-0.002&	0.031&	0.001&95&	0.13&	-0.004&	0.029&	0.001&94&	0.12&0.003&0.029& 0.001&96& 0.12 \\ \cline{2-18} 
\multicolumn{1}{c}{}                          & \multirow{4}{*}{$30\%$} & $\beta_0$     &-0.018&	0.315&	0.100&94&	1.18&	-0.014&	0.325&	0.106&94&	1.21&0.018&0.327&0.098&96&1.22  \\
\multicolumn{1}{c}{}                          &                       & $\beta_1$   & 0.014&	0.306&	0.094&94&	1.14&	0.013&	0.316&	0.100&93&	1.17&-0.009&0.304&0.093&95&1.18 \\
\multicolumn{1}{c}{}                          &                       & $\beta_2$     &-0.010&	0.117&	0.014&94	&0.46&	0.002&	0.119&	0.014&95&	0.47&-0.002&0.116&0.013&95&0.47 \\
\multicolumn{1}{c}{}                          &                       & 
$b$     & -0.004&	0.033&	0.001&96&	0.14&	-0.004&	0.032&	0.001 &95&	0.13&0.005&0.031& 0.001&96&0.13\\
\bottomrule
\end{tabular}
\footnotetext[1]{Empirical coverage rate corresponding to a 95\% credible interval for VB and MCMC}
\footnotetext[2]{Empirical coverage rate corresponding to a 95\% confidence interval for \textit{survreg}}
\end{minipage}
\end{center}
\end{table}
\end{landscape}

\subsection{Simulation results}\label{Sub.sec.results}
The numerical results from the first study are presented in Table \ref{sim.scen.res.num}. The empirical bias, SD and MSE pertaining to parameters $\beta_2$ and $b$ exhibit notable similarity among all three methods in all the scenarios. The proposed VB algorithm has smaller empirical standard deviation but similar bias, and, therefore, smaller MSE for parameters $\beta_0$ and $\beta_1$ than those of \textit{survreg} under all considered scenarios. The empirical MSEs from the VB method are approximately 5.8\% smaller for $\beta_0$ and 6.1\% smaller for $\beta_1$ than that of \textit{survreg}. This advantage is sustained even when compared to MCMC with a sample size of 300, exhibiting empirical MSE reductions of approximate 9.6\% for $\beta_0$ and 10.5\% for $\beta_1$. When the sample size is 600, the proposed VB algorithm provides similar MSEs for parameters $\beta_0$ and $\beta_1$ with MCMC in each scenario with different censoring percentages. The 95\% coverage rates yielded by all three methods exhibit remarkable consistency and closely align with the expected credible or confidence level of 0.95, ranging from 0.93 to 0.96.

Table \ref{time_comp} presents the run time required in minutes for 500 replicates for the proposed VB method and MCMC under each scenario. We see that the VB algorithm is approximately 300 times faster than MCMC.

\begin{figure}[!ht]
\centering
\begin{subfigure}{.5\textwidth}
  \centering
  \includegraphics[width = 7.8cm]{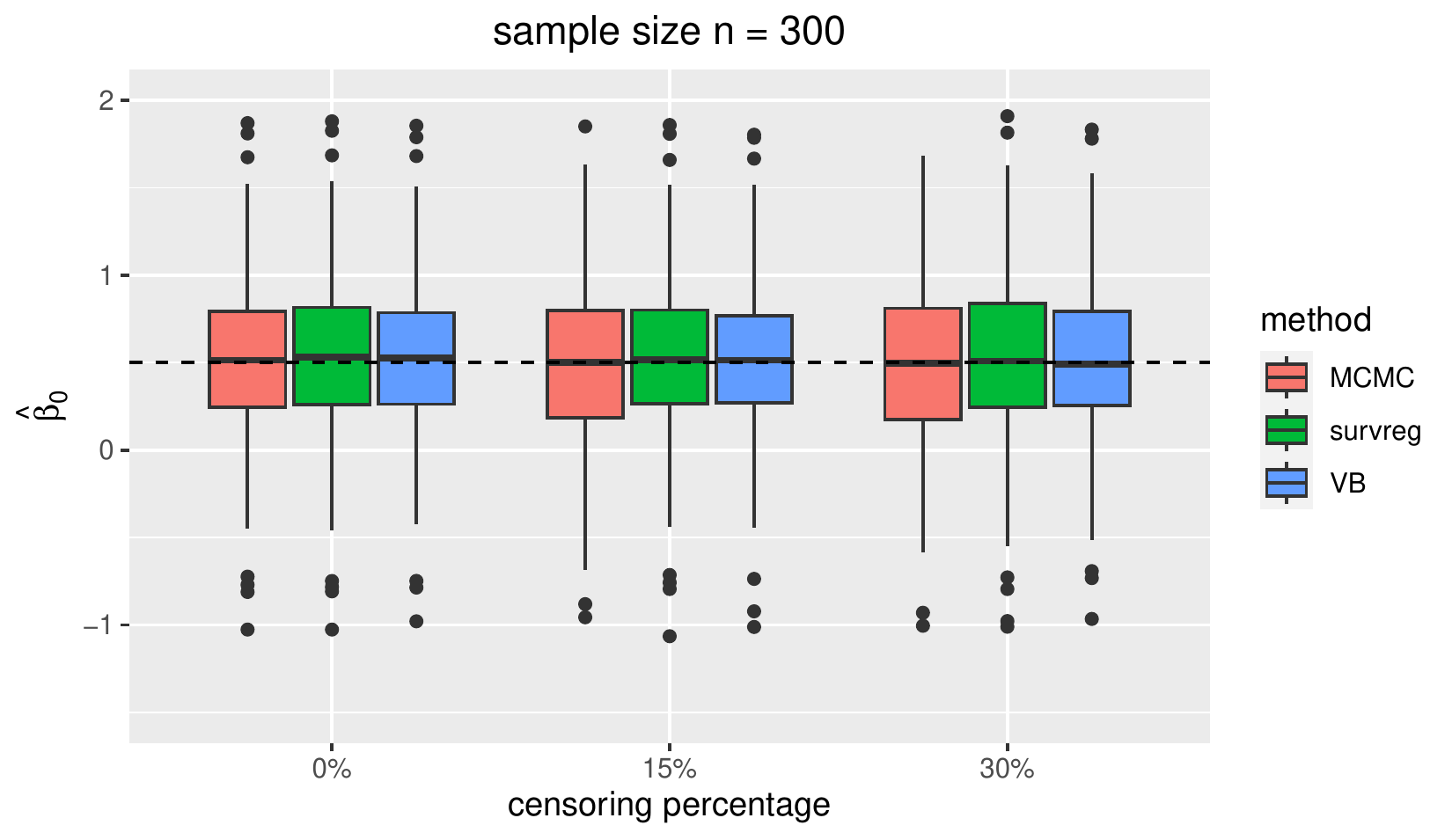}
\end{subfigure}%
\begin{subfigure}{.5\textwidth}
  \centering
  \includegraphics[width = 7.8cm]{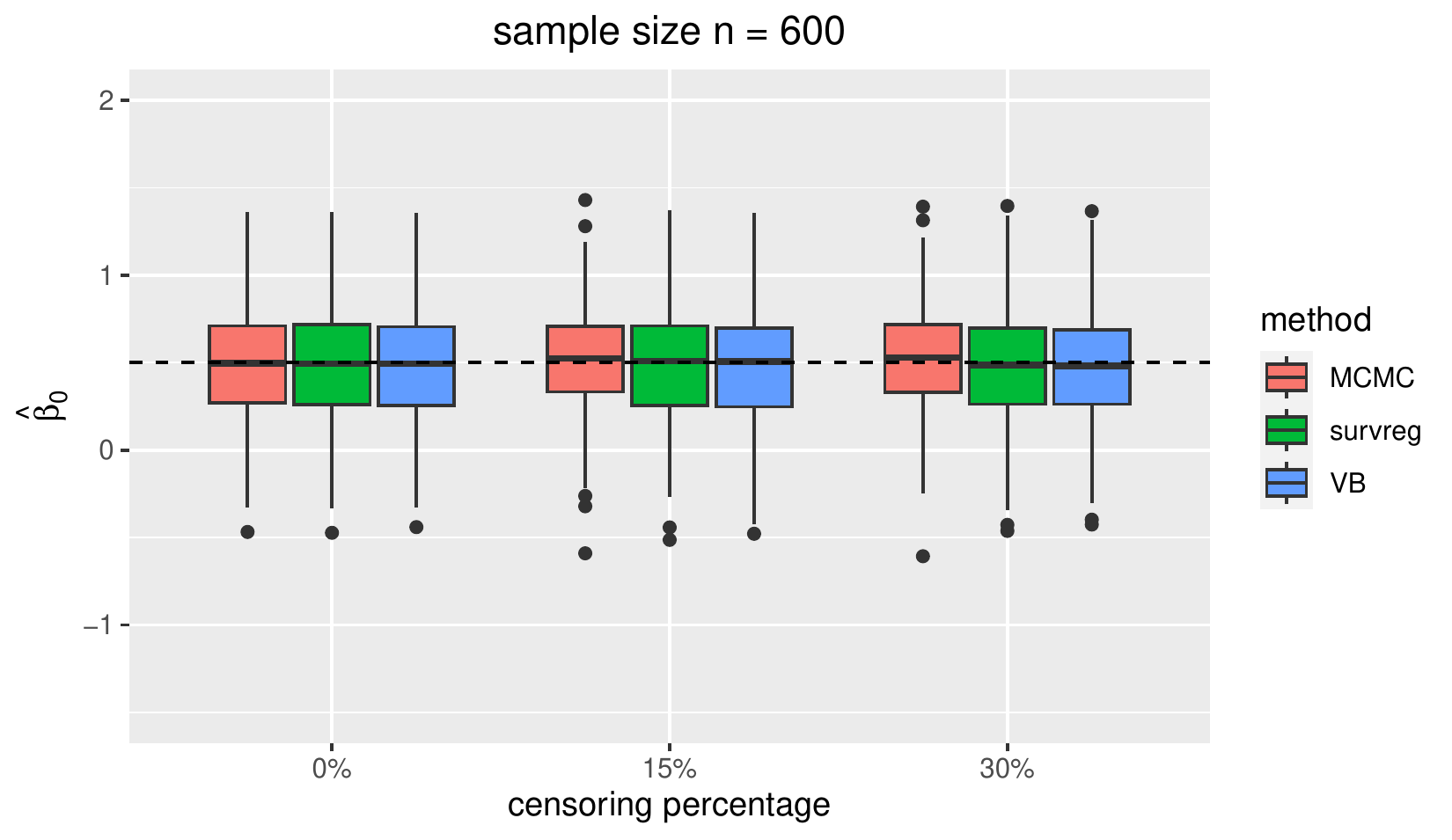}
\end{subfigure} 
\begin{subfigure}{.5\textwidth}
  \centering
  \includegraphics[width = 7.8cm]{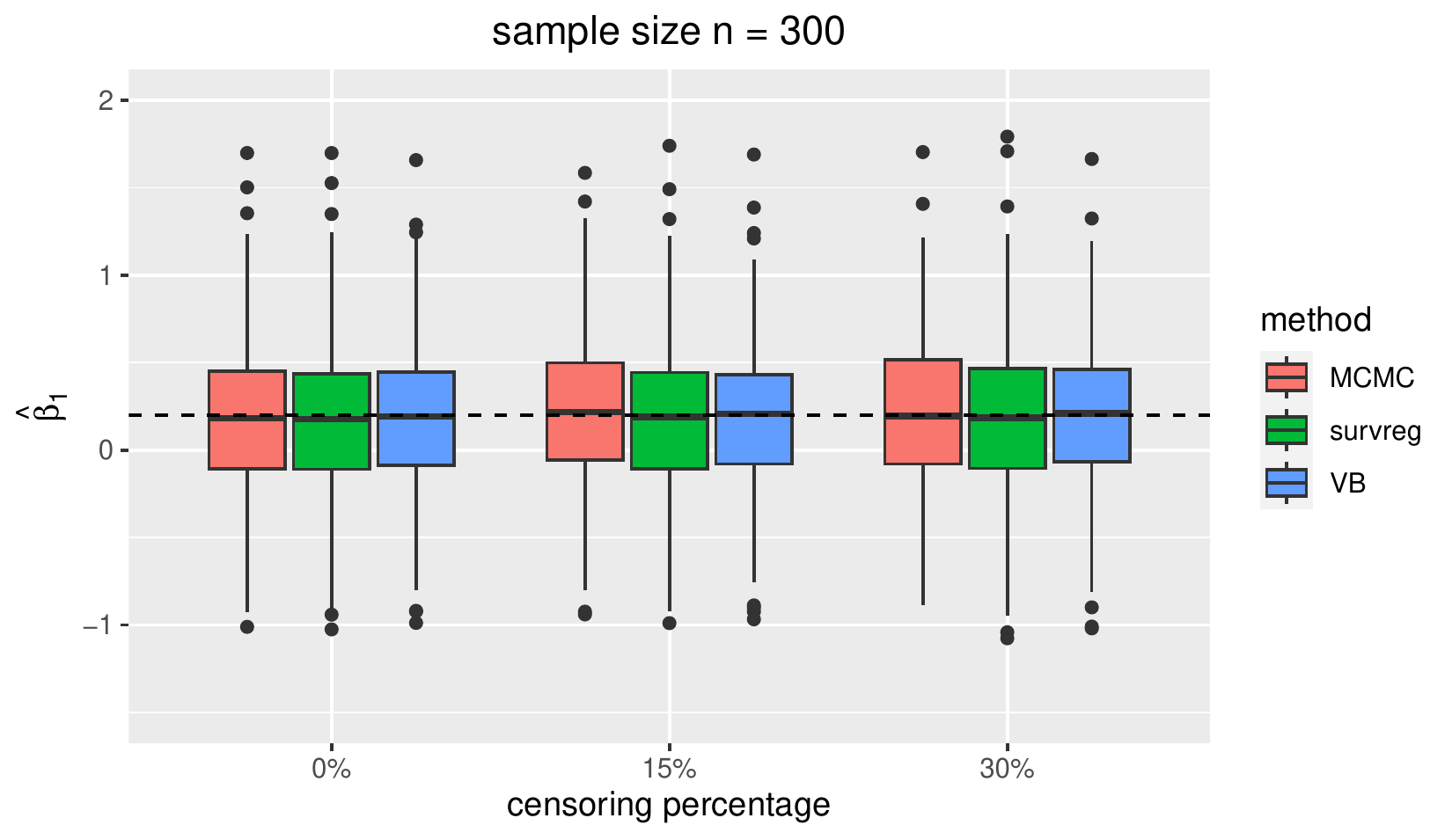}
\end{subfigure}%
\begin{subfigure}{.5\textwidth}
  \centering
  \includegraphics[width = 7.8cm]{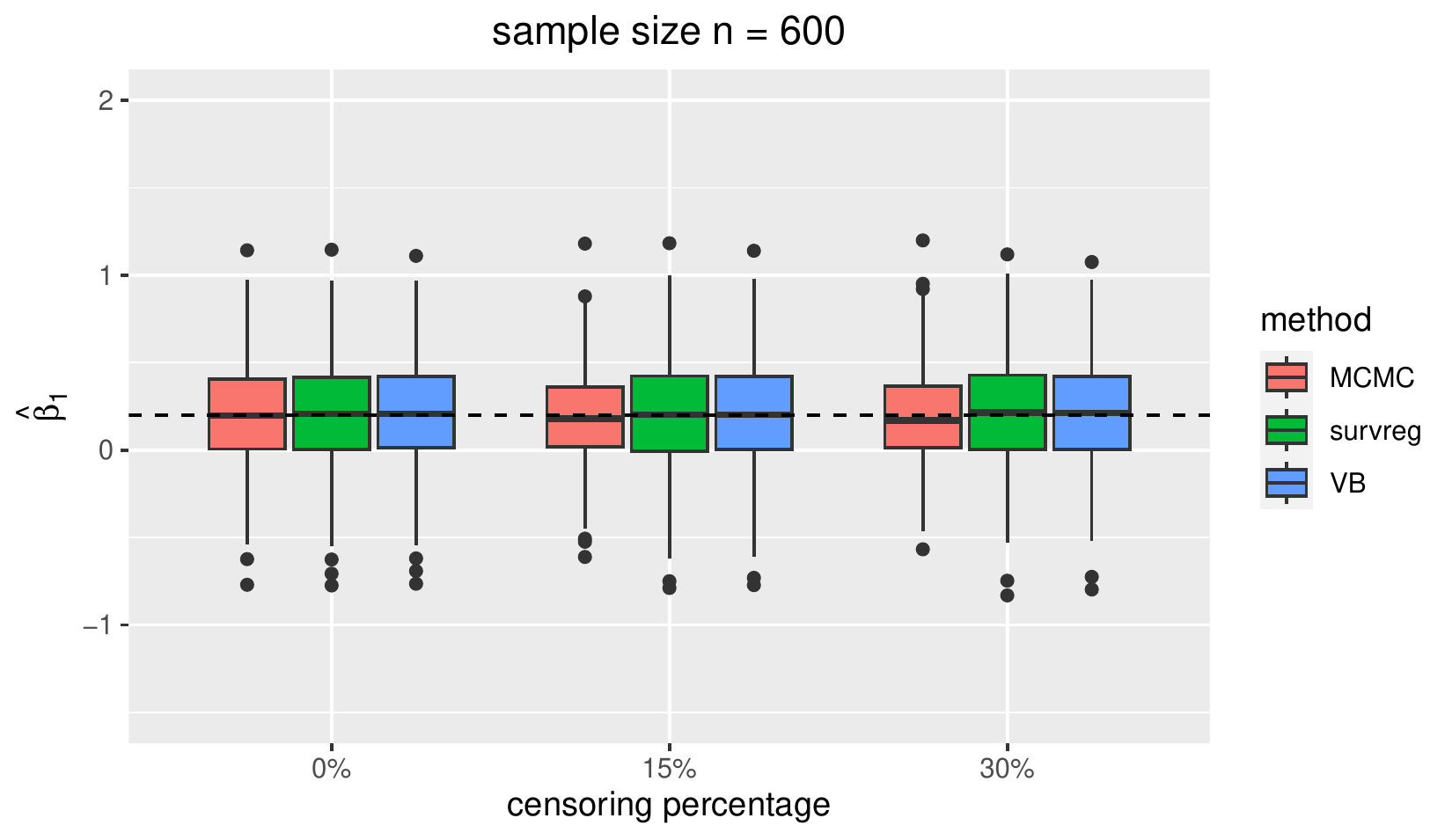}
\end{subfigure} 
\begin{subfigure}{.5\textwidth}
  \centering
  \includegraphics[width = 7.8cm]{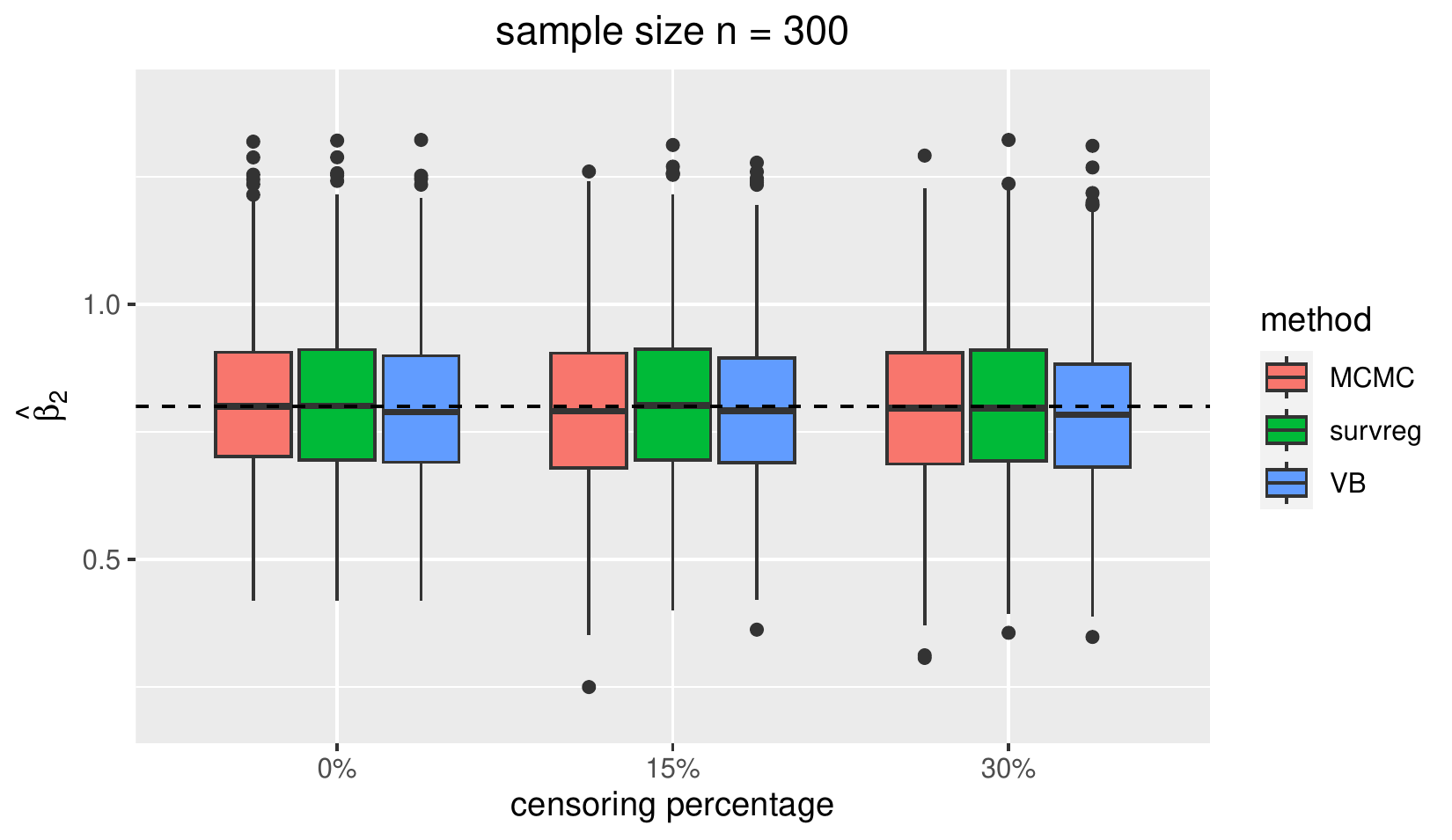}
\end{subfigure}%
\begin{subfigure}{.5\textwidth}
  \centering
  \includegraphics[width = 7.8cm]{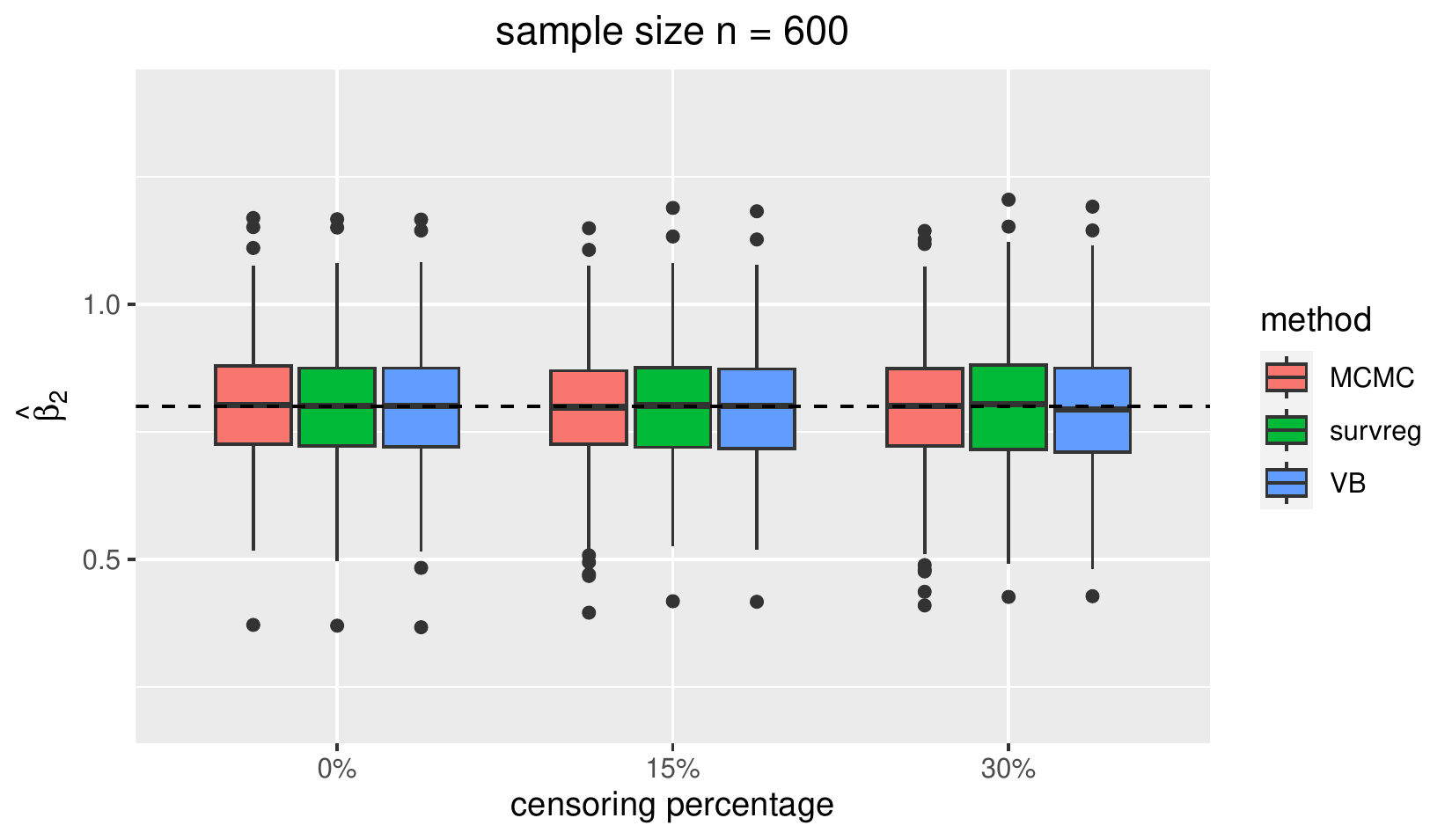}
\end{subfigure} 
\begin{subfigure}{.5\textwidth}
  \centering
  \includegraphics[width = 7.8cm]{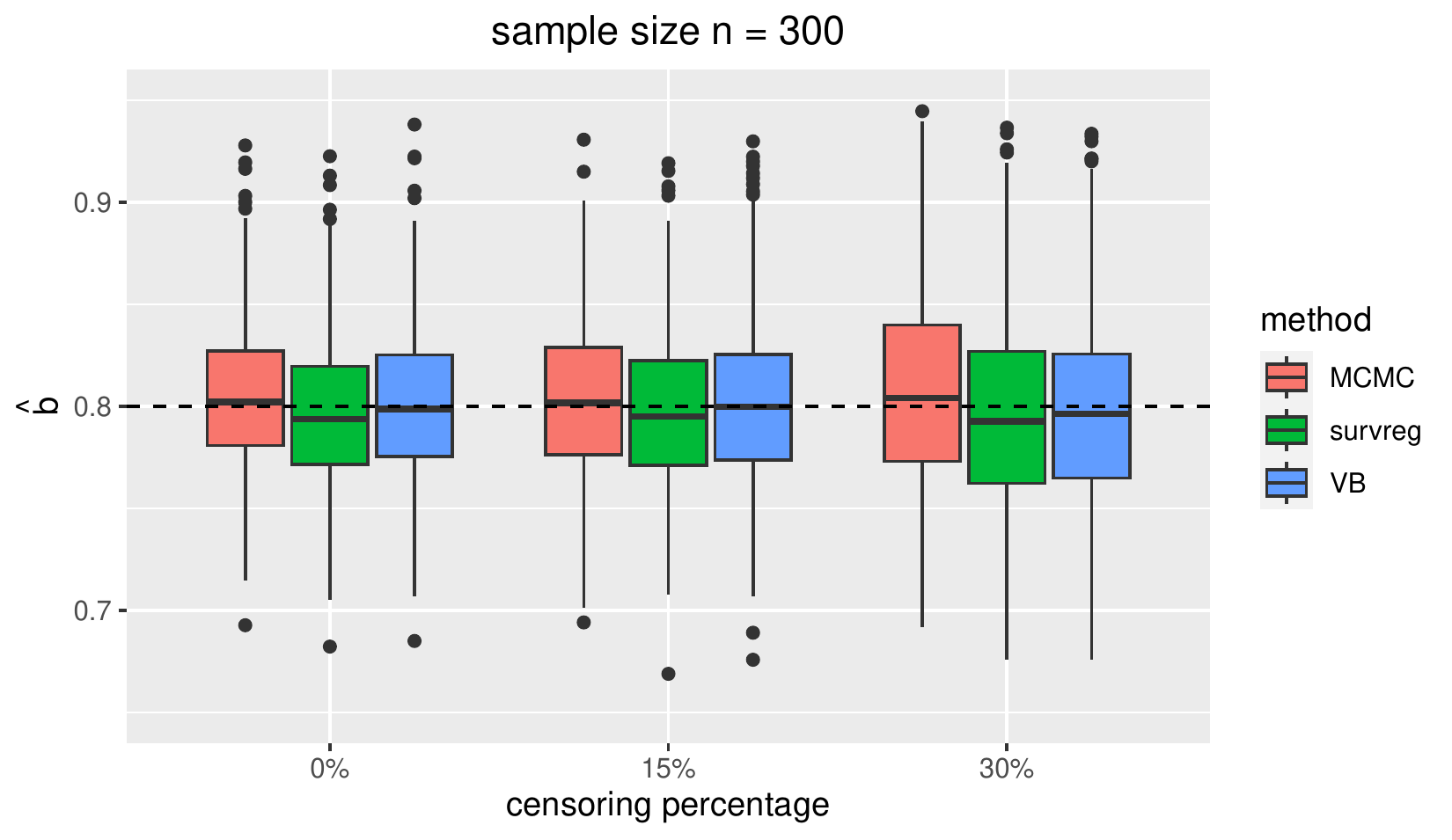}
\end{subfigure}%
\begin{subfigure}{.5\textwidth}
  \centering
  \includegraphics[width = 7.8cm]{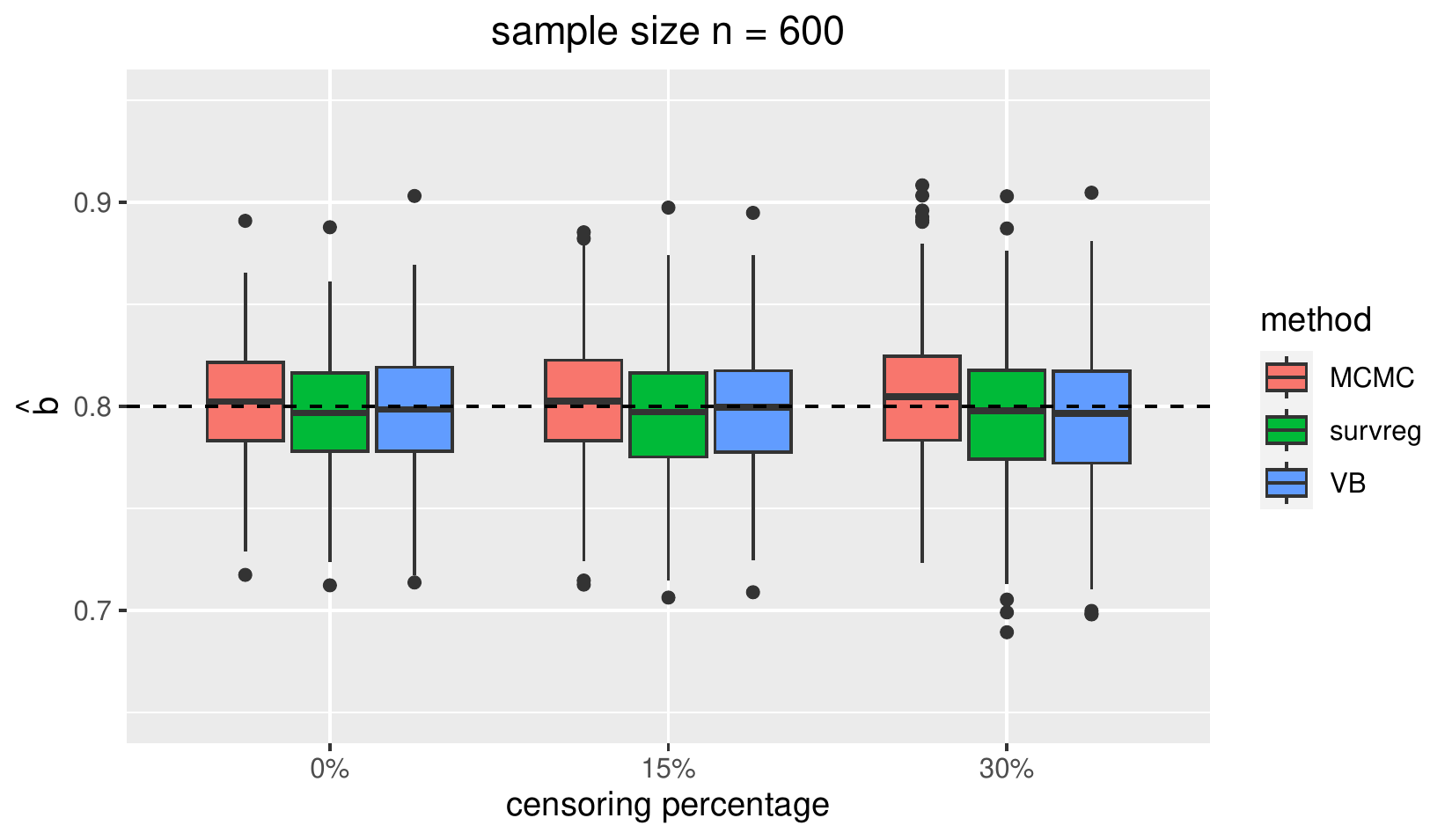}
\end{subfigure} 
\caption{\textit{\textbf{Results for simulation study one:}} A comparison of results from our VB method, the \textit{survreg} and MCMC via boxplots. The horizontal dashed line on each plot represents the true value of the corresponding parameter.}
\label{Sim.one.res.boxplots}
\end{figure}

As expected, the sample size and censoring percentage affect the MSEs. The MSE experiences an increase with higher censoring percentages and a decrease as the sample size increases. Through empirical observation, the proposed VB algorithm exhibits analogous asymptotic properties when compared to both MCMC and the likelihood-based method. To visually capture the distribution of parameter estimates across the three methods, we present side-by-side boxplots in Figure \ref{Sim.one.res.boxplots} for each parameter, considering sample sizes of $n=300$ and $n=600$.

\begin{table}[H]
\begin{center}
\caption{\textit{\textbf{Results for the first simulation study:}} Times in minutes for 500 replicates from the VB and MCMC algorithms, respectively, under scenarios with different sample sizes ($n$) and censoring percentages ($p$). The corresponding ratio of MCMC's time to VB's is also calculated and presented.}
\label{time_comp}
\begin{tabular*}{0.8\textwidth}{@{\extracolsep{\fill}}ccccccc@{\extracolsep{\fill}}}
\toprule%
$n$& \multicolumn{3}{@{}c@{}}{300} & \multicolumn{3}{@{}c@{}}{600} \\\cmidrule{2-4}\cmidrule{5-7}%
$p$ & 0\%   & 15\%   & 30\%   & 0\%   & 15\%   & 30\% \\
\midrule
VB    &   1.72  & 1.96    & 2.07    &2.81    & 3.09    &3.18     \\
MCMC  &   544.53 & 549.64   & 581.22   & 1064.22  & 1071.30   &  1109.06  \\
Ratio &   317& 280& 281& 379 &347 &349 \\ \hline
\end{tabular*}
\end{center}
\end{table}

When the sample size is small, as we considered in the second study, the MCMC provides similar estimation results as the VB method but is substantially more time-intensive in contrast to VB. We focus on the comparison between the likelihood-based \textit{survreg} method and the VB algorithm, shown in Table \ref{sim.res.two}. We observe in Table \ref{sim.res.two} that when the sample size is 30, VB consistently yields smaller MSEs across both weak and strong prior settings when contrasted with \textit{survreg}. Specifically, within the weak prior setting, VB achieves reductions in MSEs of approximately 46.6\% for $\beta_0$, 46.5\% for $\beta_1$, 8.2\% for $\beta_2$, and 42.2\% for $b$, relative to the corresponding estimates obtained via \textit{survreg}. In the strong prior setting, the reductions in MSEs are more substantial, amounting to approximately 63.4\% for $\beta_0$, 63.5\% for $\beta_1$, 15.1\% for $\beta_2$, and 39.1\% for $b$. We see that both VB and \textit{survreg} exhibit similar empirical bias for each parameter. However, estimates derived from the \textit{survreg} method are characterized by greater sample SDs, consequently leading to larger MSEs. Compared with the results in the weak prior setting, the VB method with useful prior information exhibits superior performance in estimating the regression coefficients (i.e., $\beta$'s) with smaller MSEs.

\begin{landscape}
\begin{table}[h]
\begin{center}
\begin{minipage}{1.35\textwidth}
\caption{\textit{\textbf{Results for the second simulation study:}} A comparison of numerical estimation results including the empirical Bias, sample SD, MSE, coverage rate and average interval length (Avg.L), from our VB method with two prior settings, the \textit{survreg} under a small sample size ($n=30$) with different censoring percentages ($p$).}
\label{sim.res.two}
\begin{tabular}{cccccccccccccccccc}
\toprule
\multicolumn{3}{c}{} & \multicolumn{5}{c}{VB algorithm + weak prior\footnotemark[1]}  & \multicolumn{5}{c}{VB algorithm + strong prior\footnotemark[2]} &  \multicolumn{5}{c}{\textit{survreg}} \\
\cmidrule(rl){4-8} \cmidrule(rl){9-13} \cmidrule(rl){14-18}
$n$  & $p$        &  & Bias       & SD       & MSE& Coverage\footnotemark[3]      & Avg.L      & Bias     & SD      & MSE& Coverage\footnotemark[3]      & Avg.L & Bias     & SD      & MSE& Coverage\footnotemark[4]      & Avg.L    \\ \midrule
\multicolumn{1}{c}{\multirow{12}{*}{$30$}} & \multirow{4}{*}{$0\%$}  & $\beta_0$     &-0.055&	1.102&	1.214&	95&	4.45&	-0.043&	0.917&	0.841&	95&	3.87&	-0.034&	1.473&	2.167&	91&	5.18\\
\multicolumn{1}{c}{}                          &                       & $\beta_1$     &0.047&	1.054&	1.111&	95&	4.28&	0.023&	0.877&	0.768&	96&	3.73&	0.020&	1.410&	1.984&	92&	4.97\\
\multicolumn{1}{c}{}                          &                       & $\beta_2$     &-0.081&	0.500&	0.256&	92&	1.93&	-0.055&	0.486&	0.239&	91&	1.82&	-0.067&	0.521&	0.275&	92&	1.97\\
\multicolumn{1}{c}{}                          &                       & 
$b$ & 0.010&	0.106&	0.011&	96&	0.48&	-0.032&	0.106&	0.012&	94&	0.45&	-0.036&	0.130&	0.018&	91&	0.47\\ \cline{2-18} 
\multicolumn{1}{c}{}                          & \multirow{4}{*}{$15\%$} & $\beta_0$     & -0.066&	1.101&	1.214&	95&	4.51&	-0.054&	0.912	&0.833&	95&	3.90&	-0.047&	1.501&	2.250&	92&	5.26 \\
\multicolumn{1}{c}{}                          &                       & $\beta_1$     & 0.057&	1.057&	1.118&	95&	4.34&	0.029&	0.875&	0.766&	95&	3.75&	0.033&	1.438&	2.064&	92&	5.06\\
\multicolumn{1}{c}{}                          &                       & $\beta_2$     &-0.081&	0.503&	0.259&	93&	1.96&	-0.056&	0.489&	0.242&	91	&1.83&	-0.063&	0.526&	0.280&	92&	2.00\\
\multicolumn{1}{c}{}                          &                       & 
$b$   & 0.012&	0.109&	0.012&	96&	0.51&	-0.035&	0.109&	0.013&	94&	0.48&	-0.041&	0.138&	0.021&	91&	0.50 \\ \cline{2-18} 
\multicolumn{1}{c}{}                          & \multirow{4}{*}{$30\%$} & $\beta_0$     &-0.067	&1.107&	1.228&	95&	4.62&	-0.057&0.909&	0.828&	96&	3.96&	-0.051&	1.558&	2.426&	92&	5.49 \\
\multicolumn{1}{c}{}                          &                       & $\beta_1$     &0.057&	1.056&	1.115&	95&	4.46&	0.026&	0.866&	0.749&	96&	3.82&	0.041&	1.484&	2.201&	92&	5.29\\
\multicolumn{1}{c}{}                          &                       & $\beta_2$     &-0.083&	0.513&	0.270&	94&	2.04&	-0.061&	0.492&	0.245&	93&	1.89&	-0.057&	0.545&	0.300&	92&	2.09\\
\multicolumn{1}{c}{}                          &                       & 
$b$   &0.018&	0.116&	0.014&	96&	0.55&	-0.031&	0.116&	0.014&	95&	0.51&	-0.038&	0.155&	0.025&	90&	0.56\\ 
\bottomrule
\end{tabular}
\footnotetext[1]{Weak prior setting: $\negr{\mu}_0 = (0, 0, 0)^T$, $v_0=0.1$, $\alpha_0=11$ and $\omega_0=10$}
\footnotetext[2]{Strong prior setting: $\negr{\mu}_0 = (0.3, 0.1, 1.0)^T$, $v_0=0.15$, $\alpha_0=11$ and $\omega_0=8$}
\footnotetext[3]{Empirical coverage rate corresponding to a 95\% credible interval for VB and MCMC}
\footnotetext[4]{Empirical coverage rate corresponding to a 95\% confidence interval for \textit{survreg}}
\end{minipage}
\end{center}
\end{table}
\end{landscape}

\section{Application to real data}\label{sec.real}
In this section, we apply our proposed VB algorithm in Section \ref{VBalgo} to a real data set, \textit{rhDNASE}, which is publicly available in the R package \textit{survival}. The data, first introduced in \citet{Fuchs_1994} and further analyzed in \citet{Therneau_1997}, were used to investigate the effect of recombinant human deoxyribonuclease I (rhDNase) on pulmonary function among patients with cystic fibrosis. The rhDNase can digest extracellular DNA released by leukocytes that accumulate in the airways in response to chronic bacterial infection. Therefore, administering rhDNase would reduce the incidence of exacerbation and improve lung function. Among 645 subjects, 324 were randomly assigned to the Placebo group, and the rest were assigned to the treatment group (i.e., the rhDNase group). The event time, $T$, was defined as the time until the first pulmonary exacerbation, and the follow-up period was 169 days. The forced expiratory volume (FEV) at enrollment was considered a risk factor (i.e., covariate) measuring lung capacity. In \citet{Lawless_2003}, a log-logistic AFT model was applied to this data set, and estimates were obtained by maximizing the likelihood. Model diagnostic in \citet{Lawless_2003} shows that the parametric assumption that the event time follows a log-logistic distribution was satisfied. Therefore, we want to fit the AFT regression model via our proposed VB algorithm,
\begin{eqnarray}
\log(T):= Y=\beta_0 + \beta_1 x_1 +\beta_2 x_2 + b z, \nonumber
\end{eqnarray}
where $x_1=\mathbbm{1}(\text{treatment = rhDNase})$ with $\mathbbm{1}$ being an indicator function, $x_2$ is the FEV, and $z$ follows a standard logistic distribution with a scale parameter $b$. 

Unlike simulation studies, we do not have informative priors in real data. However, we can choose priors using historical data and similar analyses on this type of data. In a similar study by \citet{Shah_1996} on the effect of rhDNase on improving lung function, researchers found that daily treatment of rhDNase could reduce the risk of developing an exacerbation by 28\%. That is, a daily administration of rhDNase can prolong the occurrence of an exacerbation by 28\%. Therefore, we can choose $\log(1.28)\approx0.25$ as the prior mean of $\beta_1$. Similarly, based on \citet{Block_2006}, the odds ratio of developing an exacerbation with one unit increase of FEV is 0.96, which indicates the corresponding time to an exacerbation occurrence increase by 4\%. Therefore, we can choose $\log(1.04)\approx0.04$ as the prior mean of $\beta_2$. For the mean of the intercept (i.e., $\beta_0$) prior distribution, we can choose the log of half of the follow-up period length, $\log(169/2) \approx 4.4$. For the precision hyperparameter $v_0$, we use a low precision, with $v_0=1$, to obtain a flat prior. For the prior of the scale parameter, we use $\alpha_0 = 501$ and $\omega_0=500$ to have a mean scale of one. To summarize, we consider the following prior distributions for the model parameters: 
\begin{itemize}
    \item $\negr{\beta} \sim N_p(\negr\mu_0, \sigma_0^2 I_{p\times p})$ with $\negr\mu_0 = (4.4, 0.25, 0.04)^T$ and $v_0=1/\sigma_0^2=1$
    \item $b \sim \text{Inverse-Gamma}\,(\alpha_0, \omega_0)$ with $\alpha_0 = 501$ and $\omega_0=500$.
\end{itemize}

\begin{figure}[!ht]
\centering
\begin{subfigure}{.5\textwidth}
  \centering
  \includegraphics[width = 7.8cm]{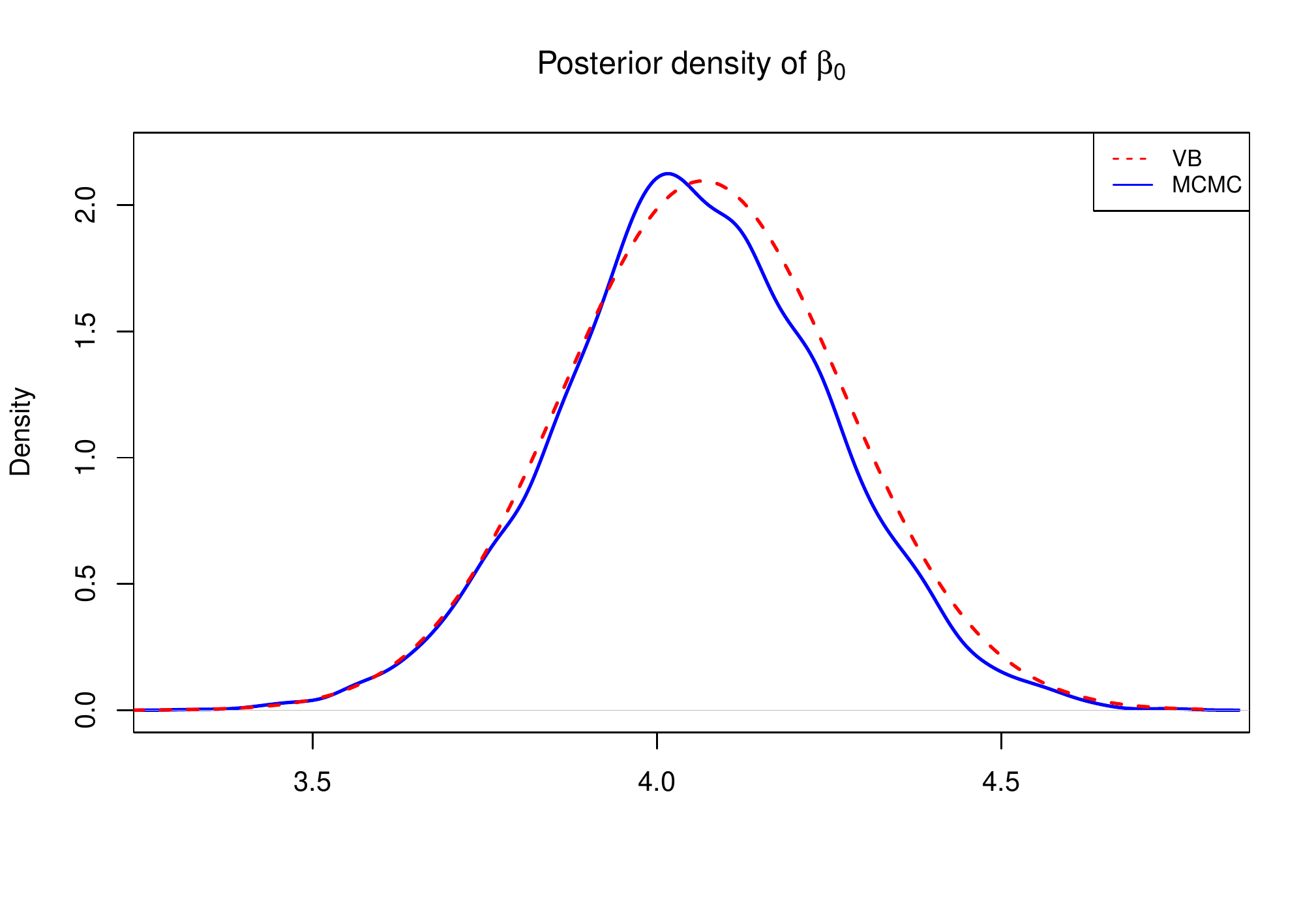}
\end{subfigure}%
\begin{subfigure}{.5\textwidth}
  \centering
  \includegraphics[width = 7.8cm]{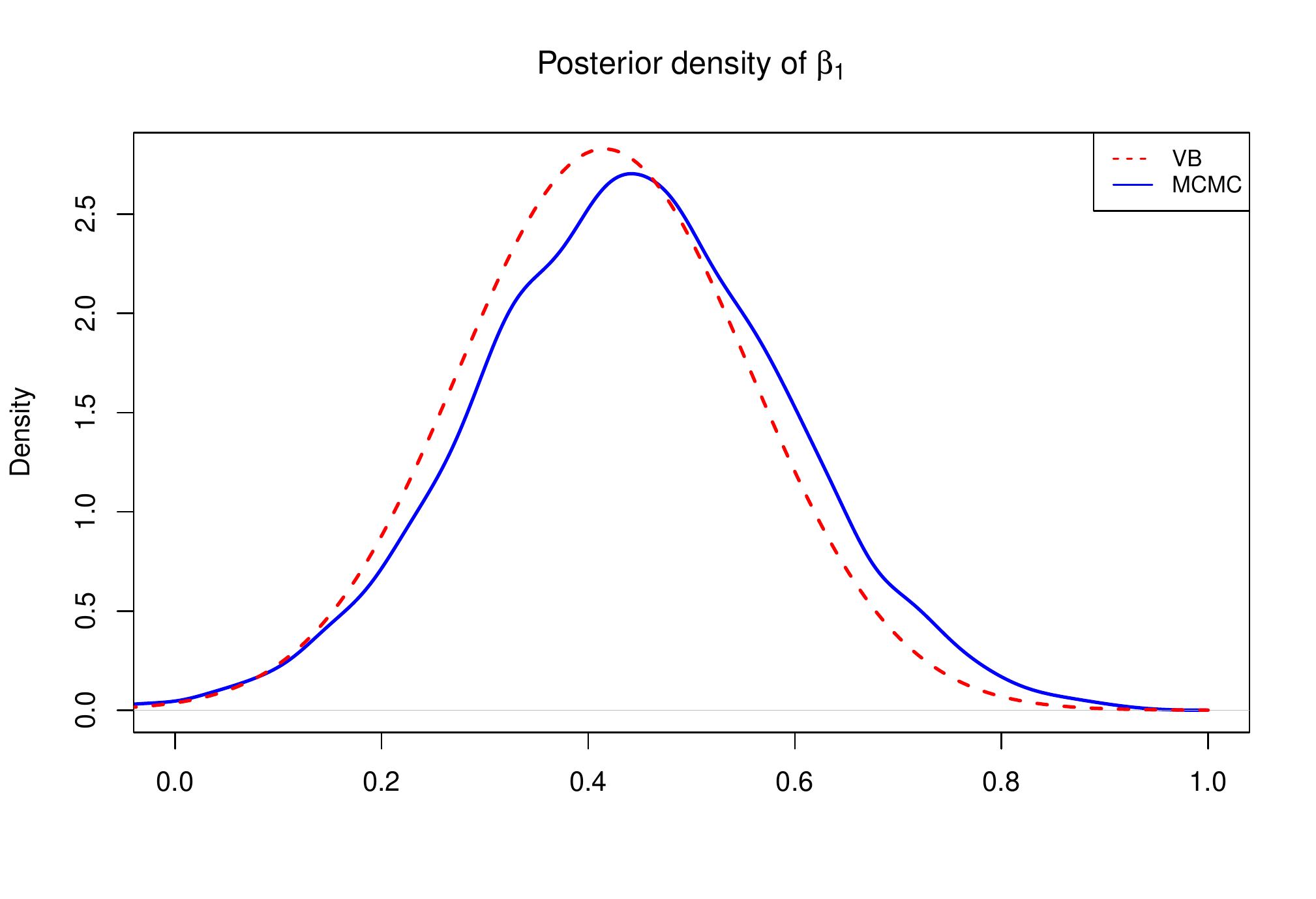}
\end{subfigure} 
\begin{subfigure}{.5\textwidth}
  \centering
  \includegraphics[width = 7.8cm]{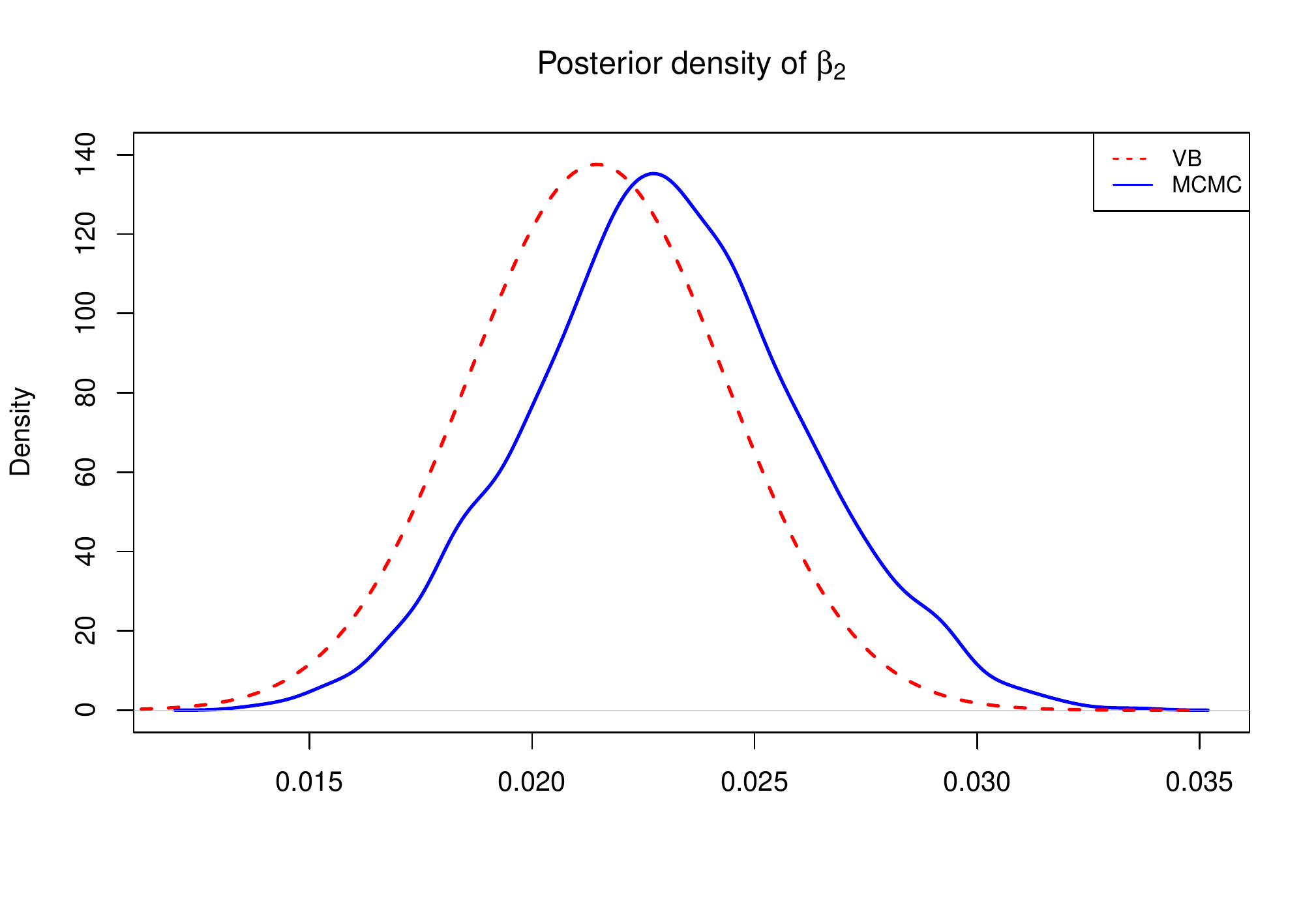}
\end{subfigure}%
\begin{subfigure}{.5\textwidth}
  \centering
  \includegraphics[width = 7.8cm]{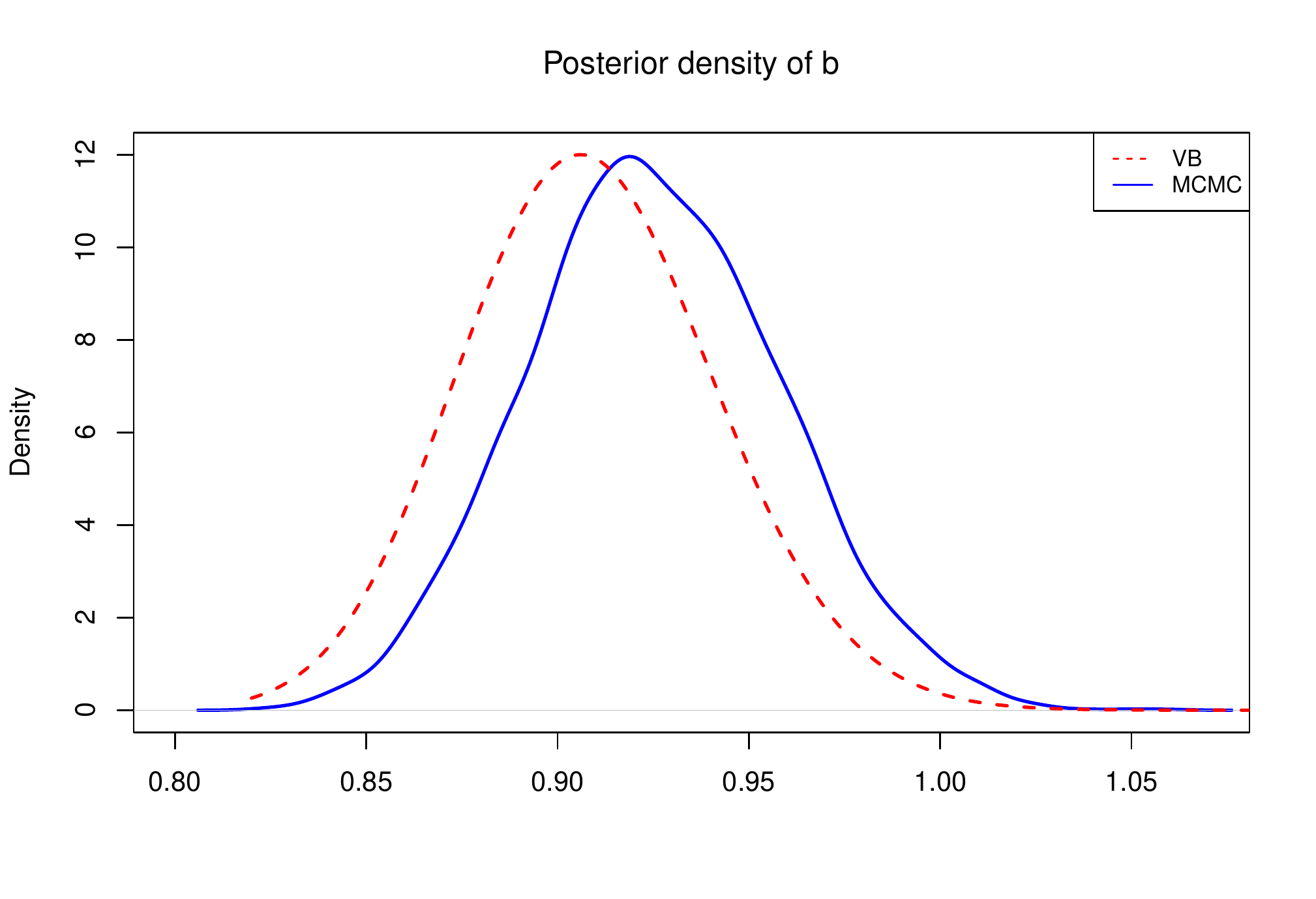}
\end{subfigure} 
\caption{\textit{\textbf{Results from analysis on rhDNASE data:}} Approximated posterior density for each parameter (dashed red line for VB and solid blue line for MCMC).}
\label{MCMC.VB}
\end{figure}

We compared the estimation results obtained using our proposed VB algorithm to those from the MCMC-based HMC algorithm and the likelihood-based survival regression, \textit{survreg}, as shown in Table \ref{tab.results.real}. The convergence of the MCMC algorithm was well assessed and checked by the trace plot and autocorrelation plot \citep{ashraf_2021}. Remarkably, all three methods exhibited a strong agreement in both point and interval estimations of each parameter. Figure \ref{MCMC.VB} depicts the approximated posterior densities of each parameter obtained from MCMC and VB, further confirming a strong agreement in the estimation of regression coefficients and the scale parameter. Notably, the computational efficiency of the proposed VB algorithm was outstanding, completing in only 0.88 seconds, whereas the MCMC method took 2.56 minutes, making it over 170 times slower than VB.

Based on the results from our VB method, the estimated coefficient of the treatment, rhDNase, is 0.416 with a 95\% credible interval of $[0.139, 0.692]$, indicating that rhDNase can significantly prolong the time to the first pulmonary exacerbation. Furthermore, the acceleration factor is $\exp(0.416) \approx 1.516$ with a 95\% credible interval of $[1.149 1.998]$ for a patient treated with rhDNase. The time to the first pulmonary exacerbation of a patient treated with rhDNase is therefore delayed by a factor of about 1.5 compared to a patient from the placebo group with the same FEV under a log-logistic AFT model. Besides, FEV is a significant risk factor on the event time, with an estimated coefficient of 0.021 (95\% credible interval $[0.016, 0.027]$). The acceleration factor of FEV is $\exp(0.021) \approx 1.021$, meaning that one unit increase in FEV would delay the event time by 2.1\% with a 95\% credible interval of $[1.6\%, 2.7\%]$. Our results from the VB algorithm highly agree with the results obtained by \textit{survreg} and the MCMC algorithm.

\begin{table}[h]
\begin{center}
\begin{minipage}{\textwidth}
\caption{\textit{Results from analysis on rhDNASE data:} Posterior means (Mean) 
with posterior standard deviations (SD), and 95\% credible intervals (95\% Cred. Int.) 
from our proposed VB algorithm and MCMC, respectively. Point estimates (Est.) with standard errors (SE), and 95\% confidence interval (95\% Conf. Int.) from \textit{survreg} in the R package \textit{survival}.}
\label{tab.results.real}
\begin{tabular*}{\textwidth}{@{\extracolsep{\fill}}cccccccccc@{\extracolsep{\fill}}}
\toprule%
& \multicolumn{3}{@{}c@{}}{VB algorithm} & \multicolumn{3}{@{}c@{}}{\textit{survreg}} & \multicolumn{3}{@{}c@{}}{MCMC}\\\cmidrule{2-4}\cmidrule{5-7}\cmidrule{8-10}%
 & Mean & SD & 95\% Cred. Int.\footnotemark[1] & Est. & SE & 95\% Conf. Int.\footnotemark[2] & Mean & SD & 95\% Cred. Int.\footnotemark[3]\\
\midrule
 $\beta_0$ & 4.113 & 0.190 & [3.740, 4.486] & 4.086 & 0.175 & [3.743, 4.429] & 4.046 & 0.198 & [3.650, 4.424] \\
 $\beta_1$ & 0.416 & 0.141 & [0.139, 0.692] & 0.402 & 0.130  & [0.146, 0.657]& 0.440 & 0.147 & [0.165, 0.737]\\
 $\beta_2$ & 0.021 & 0.003 & [0.016, 0.027]& 0.021 & 0.003 & [0.015, 0.026]& 0.023 & 0.003 & [0.017, 0.030] \\
 $b$ & 0.908 & 0.033 & [0.844, 0.974] & 0.796 & 0.045\footnotemark[4] & [0.712, 0.891] & 0.926 & 0.033 & [0.866, 0.995]  \\
\hline
\end{tabular*}
\footnotetext[1]{95\% Cred. Int.: highest density interval (HDI) was applied. Note that for a symmetric distribution, HDI is the same as the equal-tailed interval.}
\footnotetext[2]{95\% Conf. Int.: for regression coefficient estimates, the likelihood-based confidence interval was used, while for the scale estimate, the Wald-based interval was used.}
\footnotetext[3]{95\% Cred. Int.: for MCMC, we obtained the credible interval based on the percentiles of the sample from the posterior distribution.}
\footnotetext[4]{Standard error (SE) for the scale estimate is not available for \textit{survreg} in the R package, but the SE for log scale which is 0.0570, is provided. We calculated the SE for the scale estimate via Delta method.}
\end{minipage}
\end{center}
\end{table}

\section{Discussion}\label{sec.discussion}
This paper introduces a novel approach to model survival data following the log-logistic distribution as an alternative to the MCMC-based Bayesian algorithm. The study utilizes mean-field variational Bayes (VB) and applies coordinate ascent variational inference to formulate update equations within the VB framework. To achieve conjugacy under the Bayesian paradigm, the linear and quadratic piecewise approximations are embedded in the update equations for parameters. Simulation studies and the application to a real data set show that our proposed VB algorithm provides satisfactory results.

Our proposed VB approach presents several notable advantages. Similar to other Bayesian methods, our proposed VB technique accommodates the integration of prior information obtained from historical data or related studies, which is more particular in clinical research. Our VB algorithm is particularly prominent in the small sample scenario where the typical likelihood-based methods may not work well. The proposed VB algorithm also performs well under a large sample size and a weak prior setting. The proposed VB algorithm has a much lower computational cost than that of MCMC.
 
In principle, VB can be applied to the AFT regression model with other different censoring schemes, including left censored and interval censored data. However, such adaptations for a log-logistic AFT model with different censoring schemes necessitate adjustments to the likelihood function and thus to the update equation for each variational density. We anticipate that more extensive modification or a different approach altogether may be required if we consider alternative parametric distributions for survival data, for example, such as the log-normal distribution, which lacks a closed-form survival function.

To the best of our knowledge, our work stands as a pioneering effort in the application of Bayesian variational inference to model survival data via AFT regression. The piecewise polynomial approximation in the update equations is shown to work well based on the simulation studies. This approximation provides a new insight to apply Bayesian variational inference under complex models to achieve conjugacy.

\section*{Acknowledgement}
This research is supported by the Natural Sciences and Engineering Research Council of Canada (NSERC).

\printcredits

\bibliographystyle{cas-model2-names}

\bibliography{cas-refs}
\section*{Appendix}
\appendix
\section{Update equations and ELBO calculation}\label{Appe.A.VB}
In Appendix A, we derive the update equation for each component and the ELBO claculation in our model. We use $\addeq$ to denote equality up to a constant additive factor for convenience.
\subsection{VB update equations}\label{sub.sec.updates}
\noindent\textbf{(1) Update for $q^*(\negr{\beta})$}
\begin{eqnarray}
\log q^*(\negr{\beta}) &\addeq& \E_{q(b)}[\log p(\vect{D}\,\vert\,\negr{\beta},b)+\log p(\negr{\beta})] =\E_{q(b)}[\log p(\vect{D}\,\vert\,\negr{\beta},b)] + \E_{q(b)}[\log p(\negr{\beta})] \nonumber
\end{eqnarray}
where
\begin{eqnarray}
&&\E_{q(b)}[\log p(\vect{D}\,\vert\,\negr{\beta},b)] \nonumber \\
 &=& \E_{q(b)}\Bigg[-r\log b+\sum_{i=1}^n \Bigg(\delta_i \frac{y_i-\vect{X}_i^T\negr{\beta}}{b}  -(1+\delta_i)\log\Big(1+\exp\big(\frac{y_i-\vect{X}_i^T \negr{\beta}}{b}\big)\Big)\Bigg) \Bigg]\nonumber \\
 &=& -r\E_{q(b)}(\log b) + \sum_{i=1}^n \Bigg(\delta_i(y_i-\vect{X}_i^T\negr{\beta})\E_{q(b)}\Big(\frac{1}{b}\Big) -(1+\delta_i)\E_{q(b)}\bigg[\log\Big(1+\exp\big(\frac{y_i-\vect{X}_i^T\negr{\beta}}{b}\big)\Big)\bigg]\Bigg) 
 \label{update_beta_1}
\end{eqnarray}
To calculate the last expectation in (\ref{update_beta_1}) and achieve conjugacy, we then propose and apply a quadratic piecewise approximation of $\log(1+\exp(x))$ (see Equation (\ref{appro_2}) in Appendix B) to $\log\Big(1+\exp\big(\frac{y_i-\vect{X}_i^T\negr{\beta}}{b}\big)\Big)$ obtaining:
\begin{eqnarray}
&&\log\Big(1+\exp\big(\frac{y_i-\vect{X}_i^T\negr{\beta}}{b}\big)\Big) \nonumber\\
&\addeq& 0^{\nu_{i1}}\times 0.1696^{\nu_{i2}}\times 0.5^{\nu_{i3}}\times 0.8303^{\nu_{i4}}\times 1^{{1-\sum_{j=1}^4\nu_{ij}}}\frac{y_i-\vect{X}_i^T\negr{\beta}}{b} \nonumber \\
&& +\, 0^{\nu_{i1}}\times 0.0189^{\nu_{i2}}\times 0.1138^{\nu_{i3}}\times 0.0190^{\nu_{i4}}\times 0^{{1-\sum_{j=1}^4\nu_{ij}}}\Bigg(\frac{y_i-\vect{X}_i^T\negr{\beta}}{b}\Bigg)^2, \nonumber
\end{eqnarray}
where
\begin{eqnarray}
\nu_{i1}=\begin{cases} 1  &  \text{if}\; \frac{y_i-\vect{X}_i^T\negr{\beta}}{b} \leq -5\\ 0 &  \text{otherwise} \end{cases} ,\quad\nu_{i2}=\begin{cases} 1  &  \text{if}\; -5 < \frac{y_i-\vect{X}_i^T\negr{\beta}}{b} \leq -1.7\\ 0 &  \text{otherwise} \end{cases}, \nonumber
\end{eqnarray}
\begin{eqnarray}
\nu_{i3}=\begin{cases} 1  &  \text{if}\; -1.7 < \frac{y_i-\vect{X}_i^T\negr{\beta}}{b} \leq 1.7\\ 0 &  \text{otherwise} \end{cases} \quad \text{and} \quad\nu_{i4}=\begin{cases} 1  &  \text{if}\; 1.7 < \frac{y_i-\vect{X}_i^T\negr{\beta}}{b} \leq 5\\ 0 &  \text{otherwise} \end{cases}. \nonumber
\end{eqnarray}

Let $\rho_i:=0^{\nu_{i1}}\times 0.1696^{\nu_{i2}}\times 0.5^{\nu_{i3}}\times 0.8303^{\nu_{i4}}\times 1^{{1-\sum_{j=1}^4\nu_{ij}}}$ and $\zeta_i:=0^{\nu_{i1}}\times 0.0189^{\nu_{i2}}\times 0.1138^{\nu_{i3}}\times 0.0190^{\nu_{i4}}\times 0^{{1-\sum_{j=1}^4\nu_{ij}}}$,
we obtain 
\begin{eqnarray}
\log\Big(1+\exp\big(\frac{y_i-\vect{X}_i^T\negr{\beta}}{b}\big)\Big) \addeq \rho_i\frac{y_i-\vect{X}_i^T\negr{\beta}}{b} + \zeta_i \Bigg(\frac{y_i-\vect{X}_i^T\negr{\beta}}{b}\Bigg)^2.\nonumber
\end{eqnarray}
More details about the proposed quadratic piecewise approximation can be found in Appendix B. Therefore, we can write Equation (\ref{update_beta_1}) as
\begin{eqnarray}
&&  \E_{q(b)}[\log p(\vect{D}\,\vert\,\negr{\beta},b)] \nonumber \\
 &\addeq& -r\E_{q(b)}[\log b] + \sum_{i=1}^n \Bigg(\delta_i(y_i-\vect{X}_i^T\negr{\beta})\E_{q(b)}\big(\frac{1}{b}\big) - (1+\delta_i)\E_{q(b)}\Big[\rho_i\frac{y_i-\vect{X}_i^T\negr{\beta}}{b} + \zeta_i \Big(\frac{y_i-\vect{X}_i^T\negr{\beta}}{b}\Big)^2\Big]\Bigg)  \nonumber \\
 &\addeq& \sum_{i=1}^n \Bigg(-\delta_i\vect{X}_i^T\negr{\beta}\,\E_{q(b)}\big(\frac{1}{b}\big)-(1+\delta_i)\Big(-\rho_i\vect{X}_i^T\negr{\beta}\,\E_{q(b)}\big(\frac{1}{b}\big)+\zeta_i(-2y_i\vect{X}_i^T\negr{\beta}+\negr{\beta}^T\vect{X}_i\vect{X}_i^T\negr{\beta})\E_{q(b)}\big(\frac{1}{b^2}\big)\Big)\Bigg)  \nonumber \\
 &=& \sum_{i=1}^n \Bigg(\E_{q(b)}\big(\frac{1}{b}\big)\Big(-\delta_i+(1+\delta_i)\rho_i\Big)\vect{X}_i^T+2\E_{q(b)}\big(\frac{1}{b^2}\big) (1+\delta_i)y_i\zeta_i \vect{X}_i^T\Bigg)\,\negr{\beta} \nonumber\\
 &&-\,\negr{\beta}^T \Bigg(\E_{q(b)}\big(\frac{1}{b^2}\big) \sum_{i=1}^n(1+\delta_i)\zeta_i\vect{X}_i\vect{X}_i^T\Bigg)\,\negr{\beta} 
 \label{update_beta_1_new}
\end{eqnarray}
and note that
\begin{eqnarray}
\E_{q(b)}[\log p(\negr{\beta})]  \addeq \frac{p}{2}\log v_0-\frac{1}{2}v_0(\negr{\beta}-\negr{\mu}_0)^T(\negr{\beta}-\negr{\mu}_0) \addeq -\frac{1}{2}v_0\,\big[\negr{\beta}^T\negr{\beta}-2\negr{\mu}_0^T\negr{\beta}\big] = v_0\,\negr{\mu}_0^T\negr{\beta} -\frac{1}{2}v_0\negr{\beta}^T\negr{\beta}. \label{eq:beta:2}
\end{eqnarray}
Combining Equations (\ref{update_beta_1_new}) and (\ref{eq:beta:2}), we have
\begin{eqnarray}
\log q^*(\beta) &\addeq& \Bigg[v_0\,\negr{\mu}_0^T + \sum_{i=1}^n \Bigg(\E_{q(b)}\Big(\frac{1}{b}\Big)\Big(-\delta_i+(1+\delta_i)\rho_i\Big)\vect{X}_i^T+2\E_{q(b)}\Big(\frac{1}{b^2}\Big) (1+\delta_i)\zeta_i y_i\vect{X}_i^T\Bigg)\Bigg]\,\negr{\beta} \nonumber \\
&& -\, \frac{1}{2}\negr{\beta}^T\,\bigg[v_0\textbf{I}+2\E_{q(b)}\Big(\frac{1}{b^2}\Big) \sum_{i=1}^n(1+\delta_i)\zeta_i\vect{X}_i\vect{X}_i^T\bigg]\,\negr{\beta} .\nonumber
\end{eqnarray}
Let 
\begin{eqnarray}
\Sigma^*:= \bigg[v_0\textbf{I}+2\E_{q(b)}\Big(\frac{1}{b^2}\Big) \sum_{i=1}^n(1+\delta_i)\zeta_i\vect{X}_i\vect{X}_i^T\bigg]^{-1},
\label{eq_update_sigma}
\end{eqnarray}
and 
\begin{eqnarray}
\negr{\mu}^*:=\Bigg[\Bigg\{v_0\,\negr{\mu}_0^T + \sum_{i=1}^n \Bigg(\E_{q(b)}\Big(\frac{1}{b}\Big)\Big(-\delta_i+(1+\delta_i)\rho_i\Big)\vect{X}_i^T + 2\E_{q(b)}\Big(\frac{1}{b^2}\Big) (1+\delta_i)y_i \zeta_i \vect{X}_i^T\Bigg)\Bigg\}\,\Sigma^{*}\Bigg]^{T}.
\label{eq_update_mu}
\end{eqnarray}
Then, $q^*(\negr{\beta})$ is $N(\negr{\mu}^{*}, \Sigma^*)$. Therefore, we have the conjugate multivariate normal posterior distribution of $\negr{\beta}$ after applying the piecewise approximation to $\log\Big(1+\exp\big(\frac{y_i-\vect{X}_i^T\negr{\beta}}{b}\big)\Big)$.

\vspace{0.5cm}
\noindent\textbf{(2) Update for $q^*(b)$}
\begin{eqnarray}
\log q^*(b) \addeq \E_{q(\negr{\beta})}[\log p(\vect{D}\,\vert\,\negr{\beta},b)+\log p(b)] = \E_{q(\negr{\beta})}[\log p(\vect{D}\,\vert\,\negr{\beta},b)]+\E_{q(\negr{\beta})}[\log p(b)] \nonumber
\end{eqnarray}
First, we can show that,
\begin{eqnarray}
&&  \E_{q(\negr{\beta})}[\log p(\vect{D}\,\vert\,\negr{\beta},b)] \nonumber \\
 &=& \E_{q(\negr{\beta})}\Bigg[-r\log b+\sum_{i=1}^n \Bigg(\delta_i \frac{y_i-\vect{X}_i^T\negr{\beta}}{b}-(1+\delta_i)\log\Big(1+\exp\big(\frac{y_i-\vect{X}_i^T\negr{\beta}}{b}\big)\Big)\Bigg) \Bigg]\nonumber \\
 &=& -r\log b + \sum_{i=1}^n \E_{q(\negr{\beta})} \Bigg[\delta_i\frac{y_i-\vect{X}_i^T\negr{\beta}}{b}-(1+\delta_i)\log\Big(1+\exp\big(\frac{y_i-\vect{X}_i^T\negr{\beta}}{b}\big)\Big)\Bigg]
 \label{update_b_1}
\end{eqnarray}
We then propose and apply a linear piecewise approximation of $\log(1+\exp(x))$ (see Equation (\ref{appro_1}) in Appendix B) to $\log\Big(1+\exp\big(\frac{y_i-\vect{X}_i^T\negr{\beta}}{b}\big)\Big)$ obtaining:
\begin{eqnarray}
\log\Big(1+\exp\big(\frac{y_i-\vect{X}_i^T\negr{\beta}}{b}\big)\Big) \addeq 0^{\eta_{i1}}\times 0.0426^{\eta_{i2}}\times 0.3052^{\eta_{i3}}\times 0.6950^{\eta_{i4}} \times\, 0.9574^{\eta_{i5}}
 \times  1^{1-\sum_{j=1}^5\eta_{ij}}\frac{y_i-\vect{X}_i^T\negr{\beta}}{b},\nonumber
\end{eqnarray}
where
\begin{eqnarray}
\eta_{i1}=\begin{cases} 1  &  \text{if}\; \frac{y_i-\vect{X}_i^T\negr{\beta}}{b} \leq -5\\ 0 &  \text{otherwise} \end{cases} ,\;\eta_{i2}=\begin{cases} 1  &  \text{if}\; -5 < \frac{y_i-\vect{X}_i^T\negr{\beta}}{b} \leq -1.701\\ 0 &  \text{otherwise} \end{cases} , \;\eta_{i3}=\begin{cases} 1  &  \text{if}\; -1.701 < \frac{y_i-\vect{X}_i^T\negr{\beta}}{b} \leq 0\\ 0 &  \text{otherwise} \end{cases},\nonumber
\end{eqnarray}
\begin{eqnarray}
\eta_{i4}=\begin{cases} 1  &  \text{if}\; 0 < \frac{y_i-\vect{X}_i^T\negr{\beta}}{b} \leq 1.702\\ 0 &  \text{otherwise} \end{cases} \;\text{and} \quad \eta_{i5}=\begin{cases} 1  &  \text{if}\; 1.702 < \frac{y_i-\vect{X}_i^T\negr{\beta}}{b} \leq 5\\ 0 &  \text{otherwise} \end{cases}.\nonumber
\end{eqnarray}
Let 
\begin{eqnarray}
\varphi_i:=0^{\eta_{i1}}\times 0.0426^{\eta_{i2}}\times 0.3052^{\eta_{i3}}\times 0.6950^{\eta_{i4}}\times 0.9574^{\eta_{i5}}\times  1^{1-\sum_{j=1}^5\eta_{ij}},
\label{eq:varphi_i}
\end{eqnarray}
we obtain 
\begin{eqnarray}
\log\Big(1+\exp\big(\frac{y_i-\vect{X}_i^T\negr{\beta}}{b}\big)\Big) \addeq \varphi_i\frac{y_i-\vect{X}_i^T\negr{\beta}}{b}. \nonumber
\end{eqnarray}
More details about the proposed linear piecewise approximation can be found in Appendix B. Therefore, we can write Equation (\ref{update_b_1}) as
\begin{eqnarray}
\E_{q(\negr{\beta})}[\log p(\vect{D}\,\vert\,\negr{\beta},b)]
 &\addeq& -r\log b + \sum_{i=1}^n \E_{q(\negr{\beta})} \Bigg[\delta_i\frac{y_i-\vect{X}_i^T\negr{\beta}}{b}-(1+\delta_i)\varphi_i\frac{y_i-\vect{X}_i^T\negr{\beta}}{b}\Bigg] \nonumber \\
 &\addeq& -r\log b + \frac{1}{b}\sum_{i=1}^n \Big(\delta_i-(1+\delta_i)\varphi_i\Big)\Big(y_i-\vect{X}_i^T\E_{q(\negr{\beta})}(\negr{\beta})\Big)
 \label{update_b_1_new}
\end{eqnarray}
and note that
\begin{eqnarray}
\E_{q(\negr{\beta})}[\log p(b)] 
 \addeq -(\alpha_0+1)\log b -\frac{\omega_0}{b}
 \label{update_b_2_new}
\end{eqnarray}
Combining Equations (\ref{update_b_1_new}) and (\ref{update_b_2_new}), we have
\begin{eqnarray}
\log q^*(b) &\addeq& -(\alpha_0+r+1)\log b - \frac{1}{b}\bigg(\omega_0-\sum_{i=1}^n \Big(\delta_i-(1+\delta_i)\varphi_i\Big)\Big(y_i-\vect{X}_i^T\E_{q(\negr{\beta})}(\negr{\beta})\Big)\bigg).\nonumber
\end{eqnarray}
Let 
\begin{eqnarray}
\alpha^*=\alpha_0+r  \quad \text{and}\quad
\omega^*=\omega_0-\sum_{i=1}^n \Big(\delta_i-(1+\delta_i)\varphi_i\Big)\Big(y_i-\vect{X}_i^T\E_{q(\negr{\beta})}(\negr{\beta})\Big),
\nonumber
\end{eqnarray}
then $q^*(b)$ is Inverse-Gamma$(\alpha^*, \omega^*)$.

\subsection{ELBO calculation}\label{sec:ELBO}
Since our goal is to find $q(\cdot)$ that maximizes the ELBO, the ELBO is used as the convergence criterion of our VB algorithm, which is defined as follows:
\begin{eqnarray}
ELBO(q)=\E_{q}[\log p(\vect{D}, \negr{\beta}, b)]-\E_{q}[\log q(\negr{\beta}, b)], \nonumber
\end{eqnarray}
where
\begin{eqnarray}
\log p(\vect{D}, \negr{\beta}, b)=\log p(\vect{D}\,\vert\, \negr{\beta}, b)+\log p(\negr{\beta})+\log p(b) \;\text{and} \; \log q(\negr{\beta}, b)=\log q(\negr{\beta})+\log q(b).\nonumber
\end{eqnarray}
Let 
\begin{eqnarray}
\textit{diff}_{\negr{\beta}}=\E_{q}[\log p(\negr{\beta})]-\E_{q}[\log q(\negr{\beta})] \;\text{and} \; \textit{diff}_{b}=\E_{q}[\log p(b)]-\E_{q}[\log q(b)],\nonumber
\end{eqnarray}
then we can write the ELBO as
\begin{eqnarray}
ELBO(q)=\E_{q}[\log p(\vect{D}\,\vert\, \negr{\beta}, b)]+\textit{diff}_{\negr{\beta}}+\textit{diff}_{b}
\label{Appen:ELBO.cal}
\end{eqnarray}
We next present how to calculate each term in Equation (\ref{Appen:ELBO.cal}) with expectations taken with respect to the approximated variational distributions denoted by $q^*(\cdot)$. When calculating the first term, $\E_{q^*}[\log p(\vect{D}\,\vert\, \negr{\beta}, b)]$, we apply the linear piecewise approximation to $\log(1+\exp(x))$ again.
\begin{eqnarray}
&&\E_{q^*}[\log p(\vect{D}\,\vert\, \negr{\beta}, b)] \nonumber\\&=& \E_{q^*}\Bigg[-r\log b+\sum_{i=1}^n \Bigg(\delta_i \frac{y_i-\vect{X}_i^T\negr{\beta}}{b}-(1+\delta_i)\log\Big(1+\exp\big(\frac{y_i-\vect{X}_i^T\negr{\beta}}{b}\big)\Big)\Bigg) \Bigg]\nonumber \\
&\approx&  \E_{q^*}\Bigg[-r\log b+\sum_{i=1}^n \Bigg(\delta_i \frac{y_i-\vect{X}_i^T\negr{\beta}}{b}-(1+\delta_i)\varphi_i\frac{y_i-\vect{X}_i^T\negr{\beta}}{b}\Bigg) \Bigg]\nonumber \\
 &\addeq& -r\E_{q^*(b)}\Big(\log b\Big)+\sum_{i=1}^n \Big[\delta_i -(1+\delta_i)\varphi_i\Big]\E_{q^*(b)}\Bigg[\E_{q^*(\negr{\beta})} \frac{y_i-\vect{X}_i^T\negr{\beta}}{b}\Bigg]\nonumber \\
  &=& -r\E_{q^*(b)}\Big(\log b\Big) +\E_{q^*(b)}\Big(\frac{1}{b}\Big)\sum_{i=1}^n \big(\delta_i -(1+\delta_i)\varphi_i\big)\,\big(y_i-\vect{X}_i^T\E_{q^*(\negr{\beta})}\,(\negr{\beta})\big),\nonumber
\end{eqnarray}
where $\varphi_i$ is defined as Equation (\ref{eq:varphi_i}). Let $\negr{\Psi}$ be the digamma function defined as $\negr{\Psi}(x)=\frac{d}{dx}\log \Gamma(x)$, which can be easily calculated via numerical approximation. Then $\E_{q^*(b)}\log b$ can be calculated by $\E_{q^*(b)}\log b=\log(\omega^*)-\negr{\Psi}(\alpha^*).$ For $\textit{diff}_{\negr{\beta}}$, we derive its calculation as follows, using the fact that $\E(\vect X^T \vect X) = \mbox{trace}[\mbox{Var}(\vect X)] + \E(\vect X)^T\E(\vect X)$ where $\vect X$ is a column vector:
\begin{eqnarray}
\textit{diff}_{\negr{\beta}} &=&\E_{q^*}[\log p(\negr{\beta})] - \E_{q^*}[\log q(\negr{\beta})] \nonumber \\
&\addeq& \E_{q^*}[-\frac{1}{2}v_0(\negr{\beta}-\negr{\mu}_0)^T(\negr{\beta}-\negr{\mu}_0)] -\E_{q^*}[-\frac{1}{2}\log (\vert\Sigma^*\vert)-\frac{1}{2}(\negr{\beta}-\negr{\mu}^*)^T(\Sigma^*)^{-1}(\negr{\beta}-\negr{\mu}^*)] \nonumber \\
&\addeq& -\frac{1}{2}v_0[\text{trace}(\Sigma^*)+(\negr{\mu}^*-\negr{\mu}_0)^T(\negr{\mu}^*-\negr{\mu}_0)] + \frac{1}{2}\log (\vert\Sigma^*\vert). \nonumber
\end{eqnarray}
Note that
\begin{eqnarray}
\E_{q^*}[\frac{1}{2}(\negr{\beta}-\negr{\mu}^*)^T(\Sigma^*)^{-1}(\negr{\beta}-\negr{\mu}^*)]=\frac{p}{2}, \nonumber
\end{eqnarray}
which is always a constant at each iteration and therefore we ignore it. For $\textit{diff}_{b}$, we have
\begin{eqnarray}
\textit{diff}_{b}&=&\E_{q^*}[\log p(b)]-\E_{q^*}[\log q(b)]\nonumber \\
&\addeq& \E_{q^*}\Big[-(\alpha_0+1)\log b-\frac{\omega_0}{b}\Big] - \E_{q^*}\Big[\alpha^*\log \omega^* - \log (\Gamma(\alpha^*))-(\alpha^*+1)\log b-\frac{\omega^*}{b}\Big] \nonumber \\
&=& (\alpha^*-\alpha_0)\E_{q^*(b)}(\log b) + (\omega^*-\omega_0)\E_{q^*(b)}\big(\frac{1}{b}\big) - \alpha^*\log \omega^*. \nonumber
\end{eqnarray}
Since $\alpha^*$ does not change at each iteration, we remove $\log (\Gamma(\alpha^*))$ in the calculation of the ELBO.

\begin{figure}[!ht]
\centering
\begin{subfigure}{.5\textwidth}
  \centering
  \includegraphics[height = 7cm]{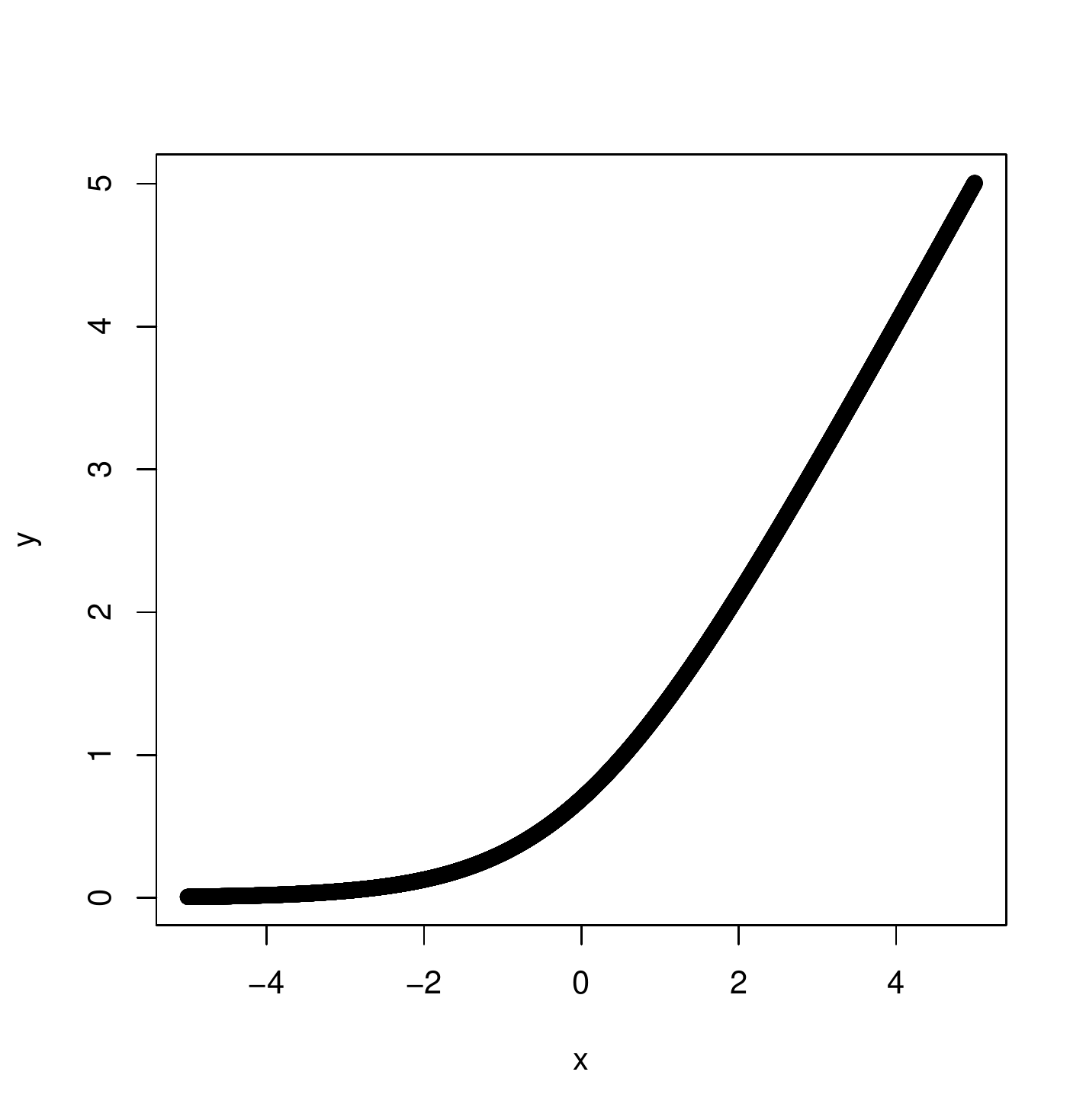}
\end{subfigure}%
\begin{subfigure}{.5\textwidth}
  \centering
  \includegraphics[height = 7cm]{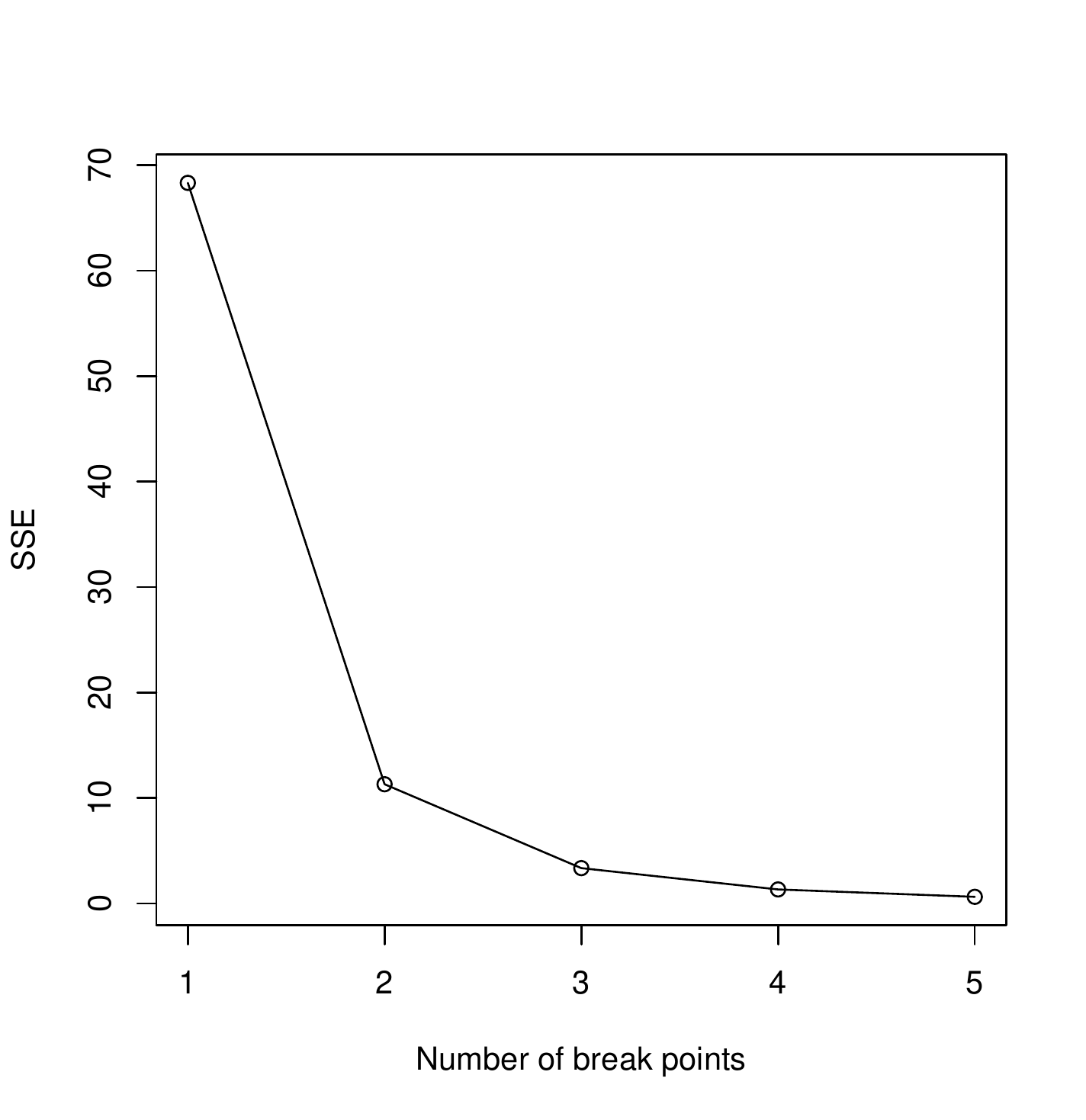}
\end{subfigure}
\caption{Left: Plot of $\log(1+\exp(x))$ versus $x$ for $x \in [-5, 5]$. Right: The plot of the sum of squared errors (SSE) versus the number of breakpoints in linear piecewise approximation via regression modelling.}
\label{piecewise_apprroximation_plots}
\end{figure}

\section{piecewise approximations of $\log(1+\exp(x))$} \label{Appe.B.piece}
This section presents the idea and details of the piecewise approximations of $\log(1+\exp(x))$. In order to have the conjugacy in our variational Bayes algorithm, we apply piecewise approximations to $\log(1+\exp(x))$, which are used in Section \ref{VBalgo}. We know that $\log(1+\exp(x))$ is monotonically increasing in $(-\infty, \infty)$, and when $x$ is approaching $-\infty$, $\log(1+\exp(x))$ approaches 0, while when $x$ is approaching $\infty$, $\log(1+\exp(x))$ approaches $x$. Furthermore, when $x \leq -5$, $\log(1+\exp(x)) \approx 0$, and when $x \geq 5$, $\log(1+\exp(x)) \approx x$ since $\log(1+\exp(-5)) = 0.0067$ and $\log(1+\exp(5)) = 5.0067$. Therefore, our goal is to find appropriate piecewise approximations of $\log(1+\exp(x))$ in $[-5, 5]$ whose plot is presented in Figure \ref{piecewise_apprroximation_plots} Left. To do this, we apply the method introduced by \citet{Muggeo_2003} implemented in R with a package called \textit{segmented} which can help find the optimal piecewise linear approximation using regression.

\begin{figure}[!ht]
\centering
\begin{subfigure}{.48\textwidth}
  \centering
  \includegraphics[height = 6.8cm]{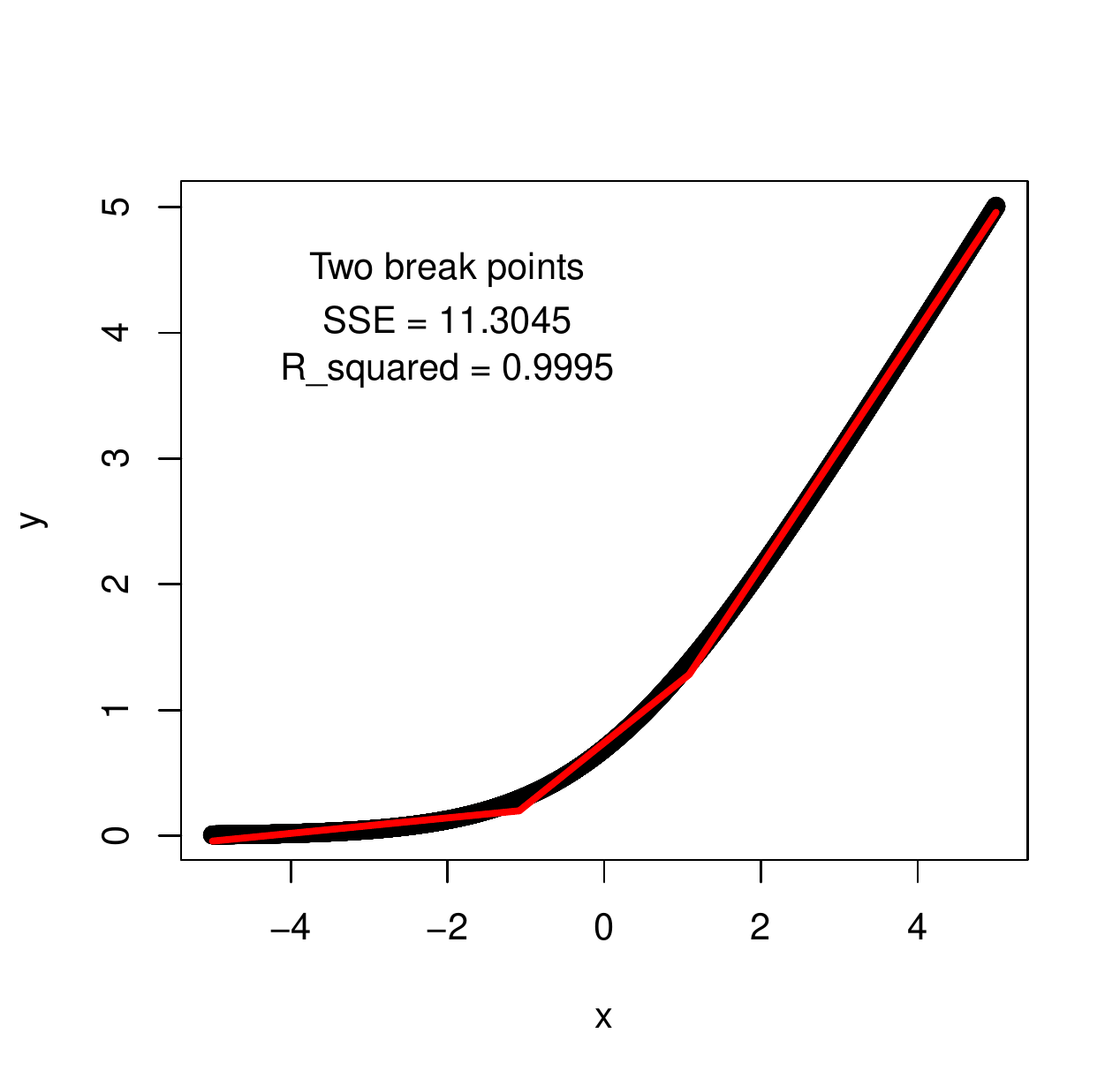}
\end{subfigure}%
\begin{subfigure}{.48\textwidth}
  \centering
  \includegraphics[height = 6.8cm]{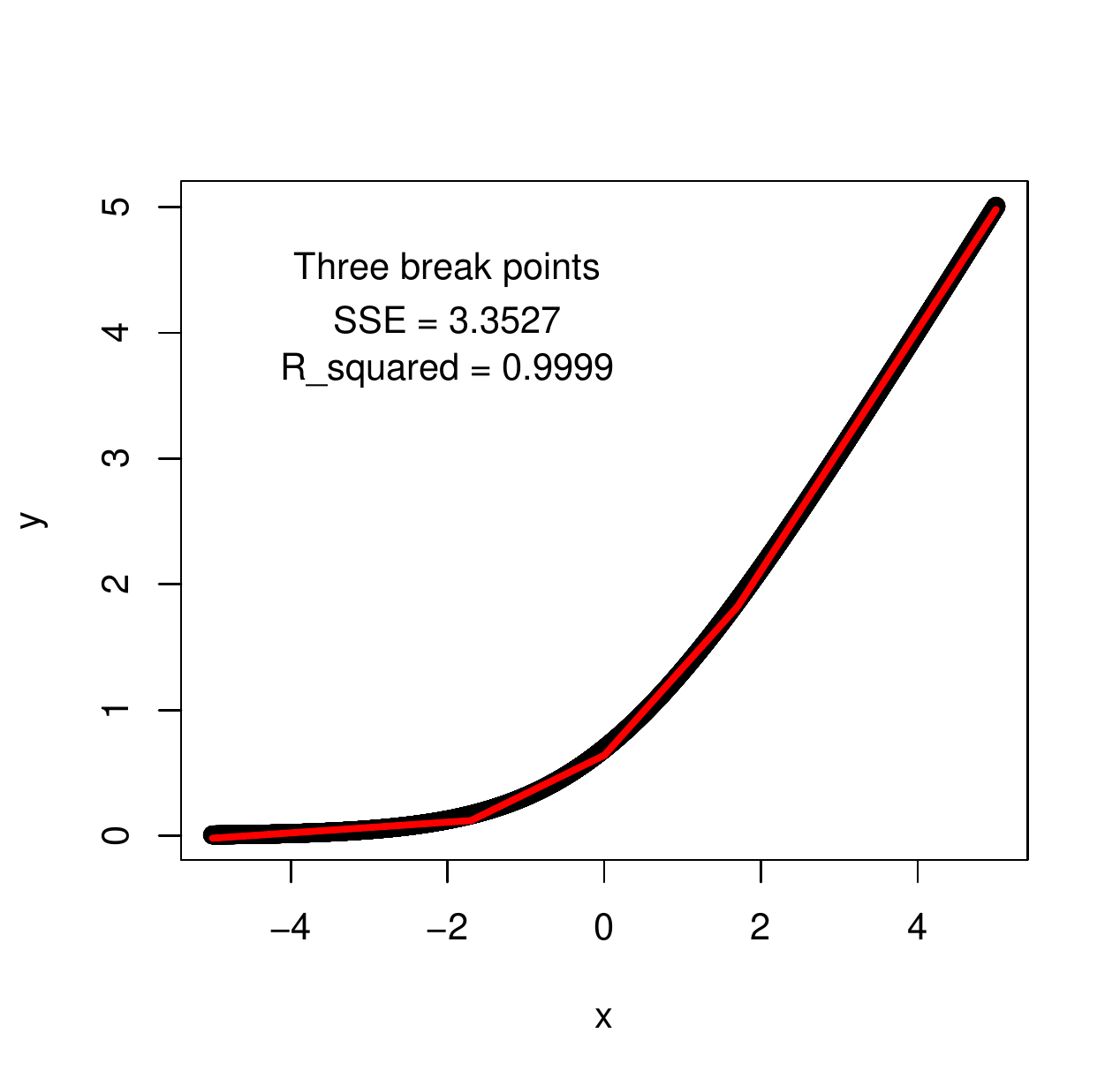}
\end{subfigure} 
\begin{subfigure}{.48\textwidth}
  \centering
  \includegraphics[height = 6.8cm]{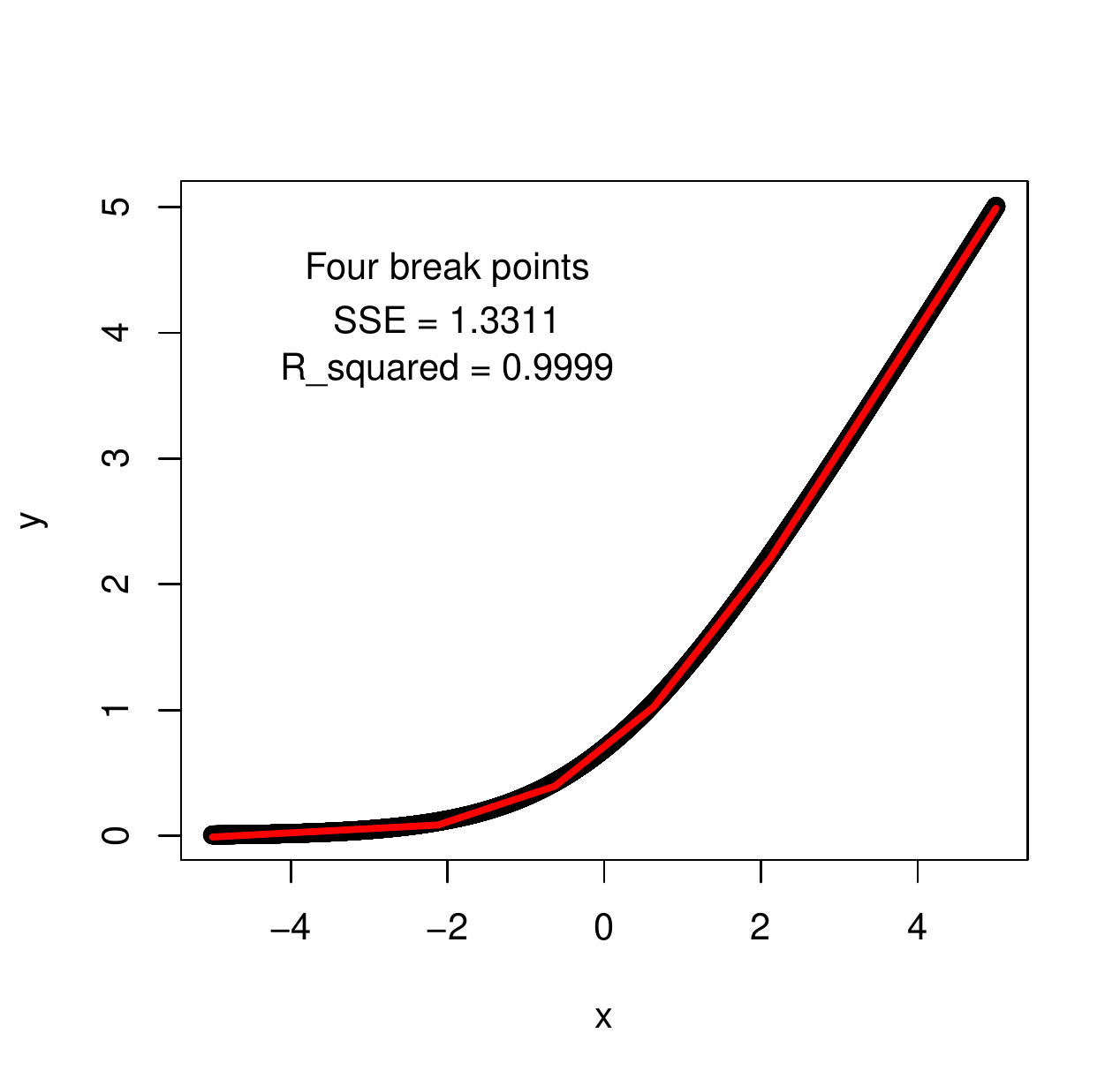}
\end{subfigure}%
\begin{subfigure}{.48\textwidth}
  \centering
  \includegraphics[height = 6.8cm]{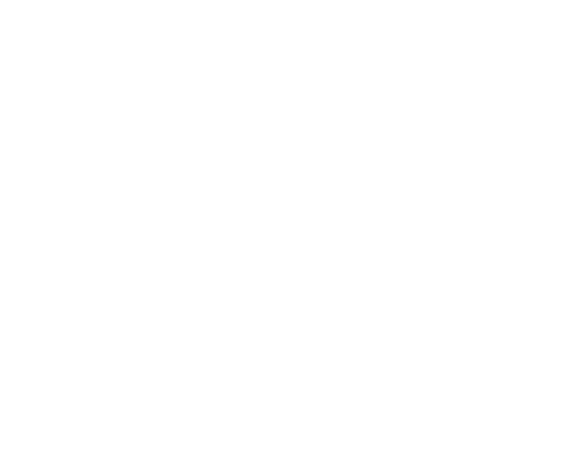}
\end{subfigure} 
\caption{A comparison of the fitted lines on the true curves using 2, 3, and 4 break points with sum of squared errors (SSE) and R squared added to the plots.}
\label{Comp_piecewise_plots}
\end{figure}

First, we generate $10, 000$ data points from $\log(1+\exp(x))$ at equally spaced grid $x_i, i=1, ..., 10000$ in $[-5, 5]$. One, two, three, four, and five breakpoints are considered, which correspond to two, three, four, five, and six pieces. The sum of squared error (SSE) is used to evaluate the performance of the fitted model on the generated data. Finally, the optimal number of breakpoints is chosen at the knee of the plot of SSE versus the number of breakpoints. From Figure \ref{piecewise_apprroximation_plots} Right, the best number of breakpoints is three with an SSE of $3.3527$ and an $R^2$ of $0.9999$. A comparison of the fitted lines on the true curves with 2, 3, and 4 breakpoints is shown in Figure \ref{Comp_piecewise_plots}. The optimal fitted model with three break points using the \textit{segmented} method proposed by \citet{Muggeo_2003} (those three optimal breakpoints are -1.701, 0, and 1.702), $\hat{f}(x)$ is
\begin{eqnarray}
\hat{f}(x) &=& 0.1938 + 0.0426x + 0.2626 (x-(-1.701))_+ \nonumber \\
 && +\,0.3898 (x-0)_+ + 0.2624(x-1.702)_+, \quad x\in [-5, 5]\nonumber
\end{eqnarray}
where $(x -a)_+ :=\text{max}(x-a, 0)$ for any $a \in (-\infty, \infty)$. 

Therefore, we can approximate $\log(1+\exp(x))$ in $(-\infty, \infty)$ by
\begin{eqnarray}
\hat{f}(x)= \widehat{\log(1+\exp(x))}=\begin{cases} 0 & \text{if}\;x \leq -5 \\ 
0.1938 + 0.0426x & \text{if}\;-5< x \leq -1.701 \\ 
0.6405 + 0.3052x & \text{if}\;-1.701< x \leq 0  \\ 
0.6405 + 0.6950x & \text{if}\; 0< x \leq 1.702 \\ 
0.1939 + 0.9574x & \text{if}\;1.702< x \leq 5  \\ 
x &  \text{if}\;5 < x 
\end{cases}
\label{appro_1}
\end{eqnarray}
We ignore the two minor jumps at $x=-5$ and $x=5$ since we focus on the approximation of the function, and manually changing the structure of the piecewise approximations will affect the optimum of the approximation.

We construct the quadratic piecewise approximation, Equation (\ref{appro_2}), based on the linear piecewise approximation. We also ignore the 
discontinuity (minor jump) at each breakpoint. The SSE using quadratic piecewise approximation is $0.1188$, and the $R^2$ of the fitted models is $1$.
\begin{eqnarray}
\hat{f}(x)= \widehat{\log(1+\exp(x))}
=\begin{cases} 0 & \text{if}\;x \leq -5 \\ 
0.3893 + 0.1696x + 0.0189 x^2 & \text{if}\;-5< x \leq -1.7 \\ 
0.6962 + 0.5000x + 0.1138 x^2 & \text{if}\;-1.7< x \leq 1.7  \\ 
0.3894 + 0.8303x + 0.0190 x^2 & \text{if}\;1.7< x \leq 5  \\ 
x &  \text{if}\;5 < x 
\end{cases}
\label{appro_2}
\end{eqnarray}

\end{document}